\documentclass[preprint2]{aastex}
\usepackage{amsmath}
\usepackage{booktabs}
\usepackage{longtable}
\usepackage{xspace}
\usepackage{hyperref}
\usepackage{comment}
\usepackage[T1]{fontenc}
\usepackage{lipsum}
\usepackage{epsfig,natbib}
\usepackage{graphicx}

\newcommand{\gcmc}{\ensuremath{\rm g\,cm^{-3}}}
\newcommand{\gcc}{\gcmc}

\newcommand{\teff}{\ensuremath{T_{\rm eff}}}
\newcommand{\logg}{\ensuremath{\log{g}}}
\newcommand{\vsini}{\ensuremath{v \sin{i}}}
\newcommand{\feh}{[Fe/H]}

\newcommand{\rsun}{\ensuremath{R_\sun}}
\newcommand{\msun}{\ensuremath{M_\sun}}

\newcommand{\rstar}{\ensuremath{R_\star}}
\newcommand{\mstar}{\ensuremath{M_\star}}

\newcommand{\rhostar}{\ensuremath{\rho_\star}}

\newcommand{\rpl}{\ensuremath{R_{\rm p}}}
\newcommand{\mpl}{\ensuremath{M_{\rm p}}}

\newcommand{\rhopl}{\ensuremath{\rho_{\rm p}}}

\newcommand{\mjup}{\ensuremath{M_{\rm J}}}

\newcommand{\rearth}{\ensuremath{R_\earth}}
\newcommand{\mearth}{\ensuremath{M_\earth}}
\newcommand{\fearth}{\ensuremath{F_\earth}}

\newcommand{\Kepler}{\textit{Kepler}}
\newcommand{\npl}{909\xspace}
\newcommand{\nstars}{355\xspace}

\newcommand{\rpratmean}{1.29}
\newcommand{\rpratmedian}{1.14}
\newcommand{\rpratstd}{0.63}

\newcommand{\pratratmean}{1.03}
\newcommand{\pratratmedian}{1.00}
\newcommand{\pratratstd}{0.27}

\begin{document}
\title{The California-Kepler Survey. V.\\
Peas in a Pod: Planets in a Kepler multi-planet System are Similar in Size and Regularly Spaced \altaffilmark{1}}
\author{Lauren M. Weiss\altaffilmark{2,3,11}}
\author{Geoffrey W. Marcy\altaffilmark{4}}
\author{Erik A. Petigura\altaffilmark{5,12}}
\author{Benjamin J. Fulton\altaffilmark{6,13}}
\author{Andrew W. Howard\altaffilmark{5}}
\author{Joshua N. Winn\altaffilmark{7}}
\author{Howard T. Isaacson\altaffilmark{4}}
\author{Timothy D. Morton\altaffilmark{7}}
\author{Lea A. Hirsch\altaffilmark{4}}
\author{Evan J. Sinukoff\altaffilmark{6,14}}
\author{Andrew Cumming\altaffilmark{2,8}}
\author{Leslie Hebb\altaffilmark{9}}
\author{Phillip A. Cargile\altaffilmark{10}}


\altaffiltext{1}{Based on observations obtained at the W.\,M.\,Keck Observatory, 
                      which is operated jointly by the University of California and the 
                      California Institute of Technology.  Keck time has been granted by 
                      the University of California, and California Institute of Technology, and the University of Hawaii.} 
\altaffiltext{2}{Institut de Recherche sur les Exoplan\`etes, Montr\'eal, QC H3T 1J4, Canada}
\altaffiltext{3}{Universit\'e de Montr\'eal, Montr\'eal, QC H3T 1J4,  Canada}
\altaffiltext{4}{University of California at Berkeley, Berkeley, CA 94720, USA}
\altaffiltext{5}{California Institute of Technology, Pasadena, CA 91125, USA}
\altaffiltext{6}{Institute for Astronomy, University of Hawaii at Manoa, Honolulu, HI 96822, USA} 
\altaffiltext{7}{Princeton University, Princeton, NJ 08544, USA}
\altaffiltext{8}{McGill University, Montr\'eal, QC H3A 0G4, Canada}
\altaffiltext{9}{Hobart and William Smith Colleges, Geneva, NY 14456, USA}
\altaffiltext{10}{Harvard-Smithsonian Center for Astrophysics, 60 Garden St, Cambridge, MA 02138, USA}
\altaffiltext{11}{Trottier Fellow}
\altaffiltext{12}{Hubble Fellow}
\altaffiltext{13}{NSF Graduate Research Fellow}
\altaffiltext{14}{Natural Sciences and Engineering Research Council of Canada Graduate Student Fellow}

\begin{abstract}
We have established precise planet radii, semi-major axes, incident stellar fluxes, and stellar masses for \npl planets in \nstars multi-planet systems discovered by \Kepler.  In this sample, we find that planets within a single multi-planet system have correlated sizes: each planet is more likely to be the size of its neighbor than a size drawn at random from the distribution of observed planet sizes.  In systems with three or more planets, the planets tend to have a regular spacing: the orbital period ratios of adjacent pairs of planets are correlated.  Furthermore, the orbital period ratios are smaller in systems with smaller planets, suggesting that the patterns in planet sizes and spacing are linked through formation and/or subsequent orbital dynamics.  Yet, we find that essentially no planets have orbital period ratios smaller than 1.2, regardless of planet size.  Using empirical mass-radius relationships, we estimate the mutual Hill separations of planet pairs.  We find that 93\% of the planet pairs are at least 10 mutual Hill radii apart, and that a spacing of $\sim20$ mutual Hill radii is most common.  We also find that when comparing planet sizes, the outer planet is larger in 65$\pm$0.4\% of cases, and the typical ratio of the outer to inner planet size is positively correlated with the temperature difference between the planets.  This could be the result of photo-evaporation.  \end{abstract}



\section{Introduction} 
Multi-planet systems provide a fossil record of the physics that drive planet formation.  The \Kepler\ Mission \citep{Borucki2010} has enabled detailed statistics of hundreds of coplanar multi-planet systems \citep{Latham2011,Lissauer2011_multis,Lissauer2012,Fabrycky2014,Lissauer2014,Rowe2014}.  In the \Kepler\ multi-planet systems, multiple planets transit the star, resulting in measured orbital periods and planet-to-star radius ratios for each transiting planet.  The observed and statistically inferred orbital properties in multi-planet systems have been used to deduce possible planet formation histories \citep{Fang2012, Hansen2013, Steffen2015, Malhotra2015, Pu&Wu2015, Ballard2016, Xie2016}.

Until recently, the stellar properties in the population of \Kepler\ multi-planet systems were poorly understood.  The majority of these stars had only photometric characterization via the Kepler Input Catalog \citep[KIC,][]{Brown2011}.  The uncertainties inherent in broad passband stellar characterization resulted in uncertainties of 16\% in the stellar masses and 42\% in the stellar radii, on average \citep{Mullally2015,Johnson2017}.

With the goal of clarifying the stellar and planetary properties of \Kepler's multi-planet systems, the California Kepler Survey (CKS) determined precise stellar and planetary properties for \nstars \Kepler\ multi-planet systems containing \npl transiting planets.  \citet[][CKS I]{Petigura2017} presented the host star spectra and their observational properties effective temperature (\teff), surface gravity (\logg), metallicity (\feh), and projected stellar rotation velocity (\vsini).  These observed quantities were converted to physical stellar parameters stellar mass (\mstar), stellar radius (\rstar), and age using stellar evolutionary models \citep[CKS II]{Johnson2017}.  The improved stellar characterization results in a median uncertainty of 5\% in the stellar mass and 10\% in the stellar radius.  

With improved stellar parameters, it is possible to improve the characterization of planets as well.  In CKS II, the updated stellar parameters were used to compute planetary radii (\rpl), semi-major axes ($a$), and equilibrium temperatures ($T_\mathrm{eq}$) for the planets orbiting these stars.  The improved stellar and planetary parameters enable a more accurate and precise characterization of the multi-planet systems than was previously available. 

In this paper, we examine several properties of Kepler's multi-planet systems that are clarified by the improved stellar parameters.  In section \ref{sec:sample} we describe how the multi-planet systems analyzed herein were selected.  In section \ref{sec:sizes} we show that the planetary sizes are related within multi-planet systems.  In section \ref{sec:spacing} we show that the period ratios between adjacent planets are related within multi-planet systems.  In section \ref{sec:underpinnings}, we explore the relation between these patterns and search for underlying physics.  In particular, we show that planet size and planet spacing are correlated.  Using the updated planet radii and semi-major axes, we employ empirical mass-radius relationships to compute the pairwise mutual Hill separations for the multis.  We also explore a correlation between the ratio of planet sizes and their equlibirium temperatures.  We conclude in section \ref{sec:conclusion}.


\section{The Sample}
\label{sec:sample}
The initial set of CKS systems with multiple transiting planet candidates consists of 469 stars with at least two transit-like signals, and a total of 1215 transit-like signals that were at one time flagged as Kepler Objects of Interest (KOIs).  From these, we discarded the known false positives, removing 59 non-planetary signals as determined in CKS I.  We then discarded stars that were diluted by at least 5\% by a second star in the \Kepler\ aperture \citep[as determined in the stellar companion catalog of][]{Furlan2017}, removing 30 stars hosting 69 planet candidates.  We discarded planets for which \citet{Mullally2015} measured $b > 0.9$, for which the high impact parameters adversely affected our ability to determine accurate planet radii, removing 75 planet candidates.  We removed planets for which the measured signal-to-noise ratio (SNR) was less than 10 since these planets have poorly determined radii and impact parameters, removing 48 planet candidates\footnote{We also tried a more conservative cutoff of SNR $>$ 20, which did not change the results.}.  Finally, we discarded systems that have been reduced to one valid planet candidate (55 systems).  After these cuts, our sample of ``CKS multis'' contained \npl high-purity planet candidates, which we henceforth call planets, in \nstars multi-planet systems.  

\begin{figure}[htbp]
\begin{center}
\includegraphics[width=\columnwidth]{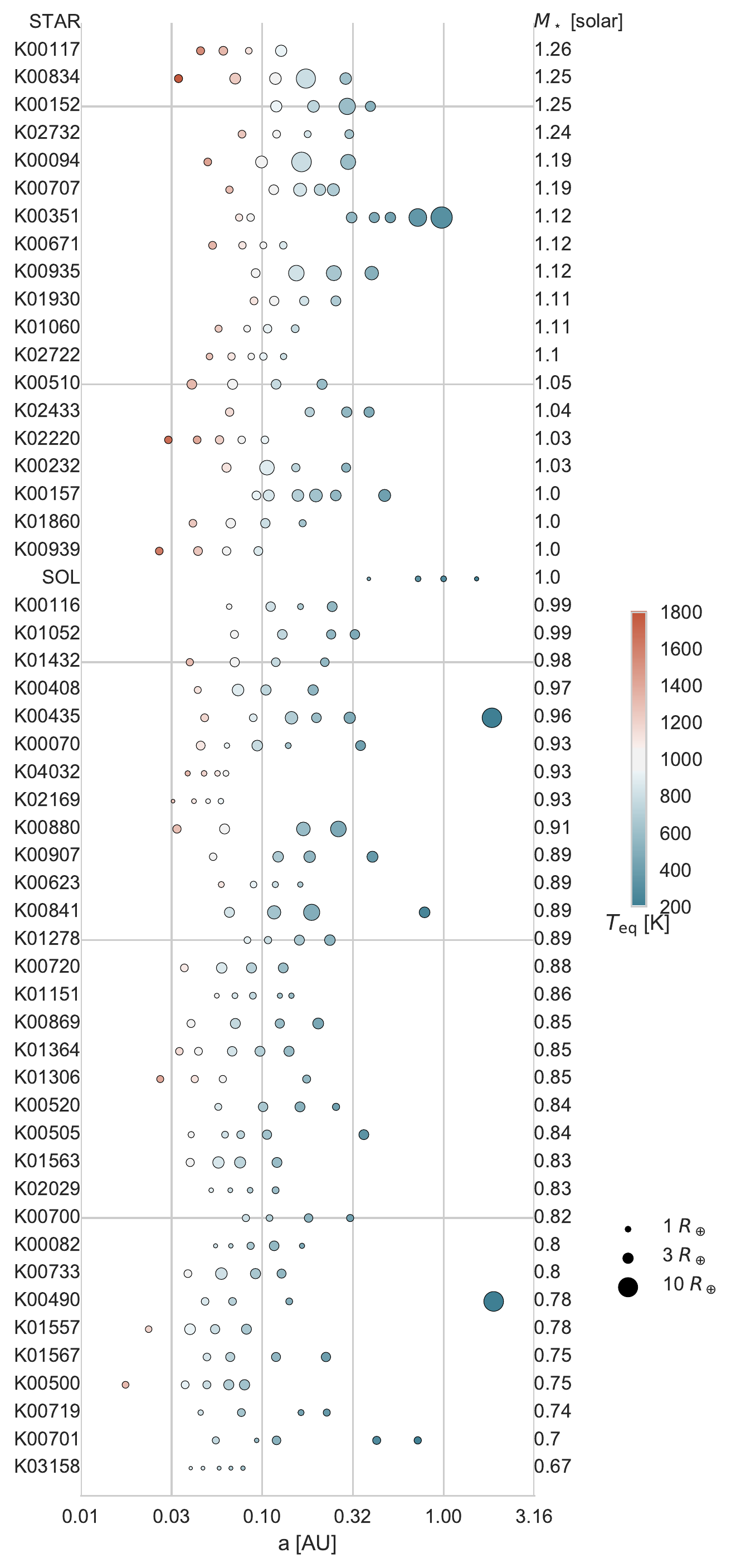}
\caption{Architectures of CKS multis with at least 4 transiting planets.  Each row corresponds to one planetary system (name on y-axis) and shows the planet semi-major axes (x-axis; note the log scale).  The point sizes correspond to the planet radii, and the point colors correspond to the equilibrium temperatures (see key to the right).  The systems are ordered by stellar mass, which is listed to the right of each system.  The inner solar system is included for comparison.}
\label{fig:architectures}
\end{center}
\end{figure}

Figure \ref{fig:architectures} shows the architectures of CKS systems with at least 4 transiting planets.  Each row corresponds to one planetary system.  The systems are ordered by stellar mass, which is listed to the right of each system.  We identify several architectural features by eye, which merit further investigation: (1) the size of one planet in a system is a good predictor of the sizes of other planets in the same system, (2) the spacing between a pair of planets in a system is a good predictor of the spacing of additional planets in that same system, (3) the smallest planets have the closest spacings, (4) when planets are not the same sizes, the outer planets are usually larger. Below, we quantitatively investigate these observations.


\section{Planets in the Same System have Similar Sizes}
\label{sec:sizes}
Is the size one planet in a given system a good predictor of the size of the next planet?  To test this, we measured the correlation between the size of a planet, $R_j$, and the size of the next planet in the system, $R_{j+1}$, in order of increasing orbital period.  To avoid detection-based asymmetries in the distribution, we only included pairs that were detectable with signal-to-noise ratio (SNR) $>$10 when their orbital positions were swapped.  The expected SNR of a planet with size $\rpl$ and orbital period $P$ orbiting a star with bulk density $\rhostar$, radius \rstar, and 6-hour Combined Differential Photometric Precision \citep[$\mathrm{CDPP_{6h}}$, a measure of the stellar photometric noise over 6 hours,][]{Christiansen2012} is:

\begin{equation}
\mathrm{SNR} = \frac{(\rpl/\rstar)^2 \sqrt{3.5 \mathrm{yr}/P}}{\mathrm{CDPP_{6h}} \sqrt{6 \mathrm{hr} / T}}
\label{eqn:detectability}
\end{equation}
\begin{equation}
T = 13\mathrm{hr}~(P/1 \mathrm{yr})^{1/3} (\rhostar/\rho_\odot)^{-1/3}
\end{equation}
If the smaller planet, when placed at the larger orbital period, produced SNR $> 10$, the planets were included in our sample of adjacent pairs.  Of the 554 original pairs of planets, 504 passed the swapping criterion.  We then tested whether a correlation was present in these pairs of planets.

\begin{figure}[htbp] 
   \begin{center}
   \includegraphics[width=\columnwidth]{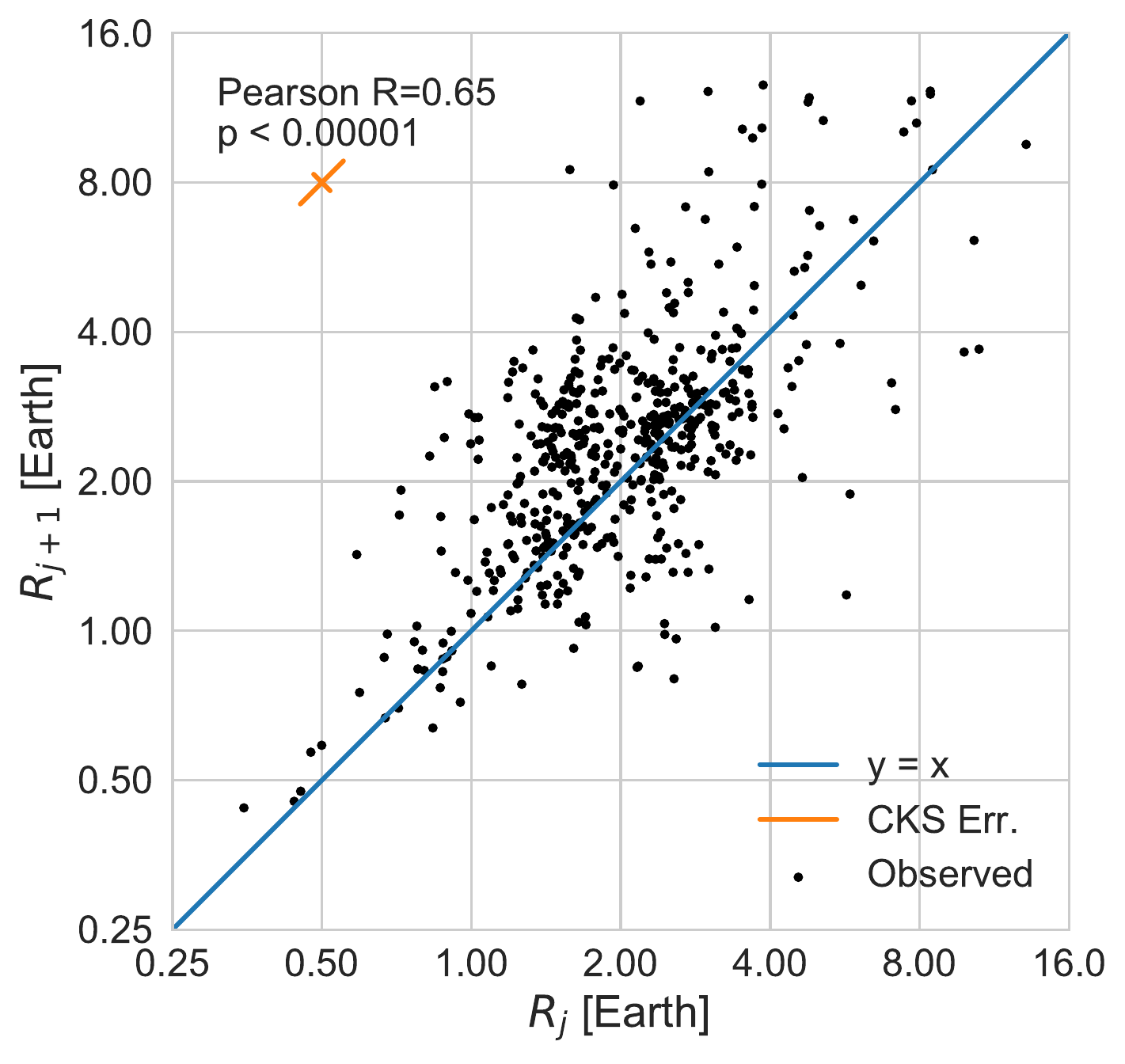} 
   \caption{The radius of a planet, $R_j$, vs.~the radius of the next planet out, $R_{j+1}$, in the CKS multi-planet systems.  There is a large and significant correlation (Pearson-R=0.62, p $< 10^{-5}$) between the size of a planet and its neighbor.  Although the outer planet radius is correlated with the inner planet radius, 65.4$\pm0.4$\% of the points sit above the $y=x$ line.  Uncertainties in the stellar radii lead to an 11\% uncertainty in the position of each point along the $y=x$ axis; uncertainties in transit depths lead to a $\sqrt{2}\times4\%$ uncertainty in the position of each point along the $y=-x$ axis (see orange cross).}
   \label{fig:rpcorr}
   \end{center}
\end{figure}

Using the Pearson-R correlation test, we find that there is a large ($r=0.65$) and significant ($p < 10^{-5}$) correlation between the sizes of adjacent planets (Figure \ref{fig:rpcorr}).  \textit{A planet in a multi-planet system is likely to be the size of its neighbor.}

A correlation in adjacent planet sizes can arise from either astrophysics or from detection biases.  One might imagine that the correlation is driven by the tail end of small planets ($\rpl < 1~\rearth$).  Planets of this size are only detectable around $\sim10\%$ of the stars in our sample ($\sim30$ stars), and the smaller planets are even less detectable.  If we restrict our sample to planets with $\rpl > 1~\rearth$, the Pearson-R correlation has $r=0.53$ and $p < 10^{-5}$, meaning that the correlation between the sizes of adjacent planets larger than 1.0~\rearth\ is still strong and significant.  Furthermore, although planets smaller than 1.0~\rearth\ would be hard to detect around other stars, it would be easy to detect larger transiting planets in the same systems, if they existed.

\begin{figure}[htbp] 
   \begin{center}
\includegraphics[width=\columnwidth]{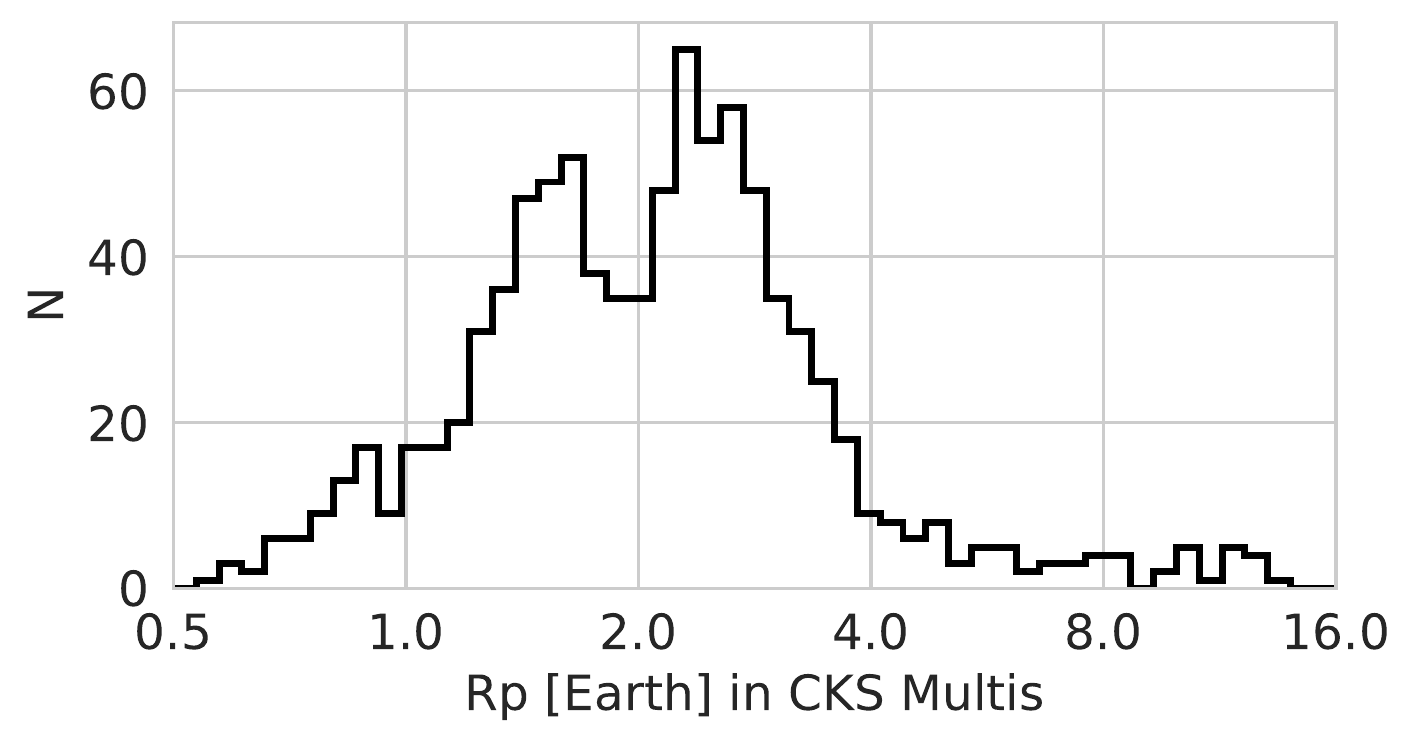}
   \caption{The radius distribution of the 909 planets in the CKS multis.  The Fulton Gap at 2~\rearth\ is visible \citep{Fulton2017}.  Note that planets larger than 4~\rearth\ constitute only 7.8\% of the distribution, and planets larger than 8~\rearth\ account for just 2.4\% of the distribution.  The majority of the patterns presented herein are driven by the sub-Neptune sized planets, not giant planets.  Planets were drawn at random, with replacement, from this distribution to construct bootstrap trials investigating planet radii like the one shown in Figure \ref{fig:rp_bootstrap}.}
   \label{fig:rp_dist}
   \end{center}
\end{figure}

\begin{figure}[htbp] 
   \begin{center}
   \includegraphics[width=\columnwidth]{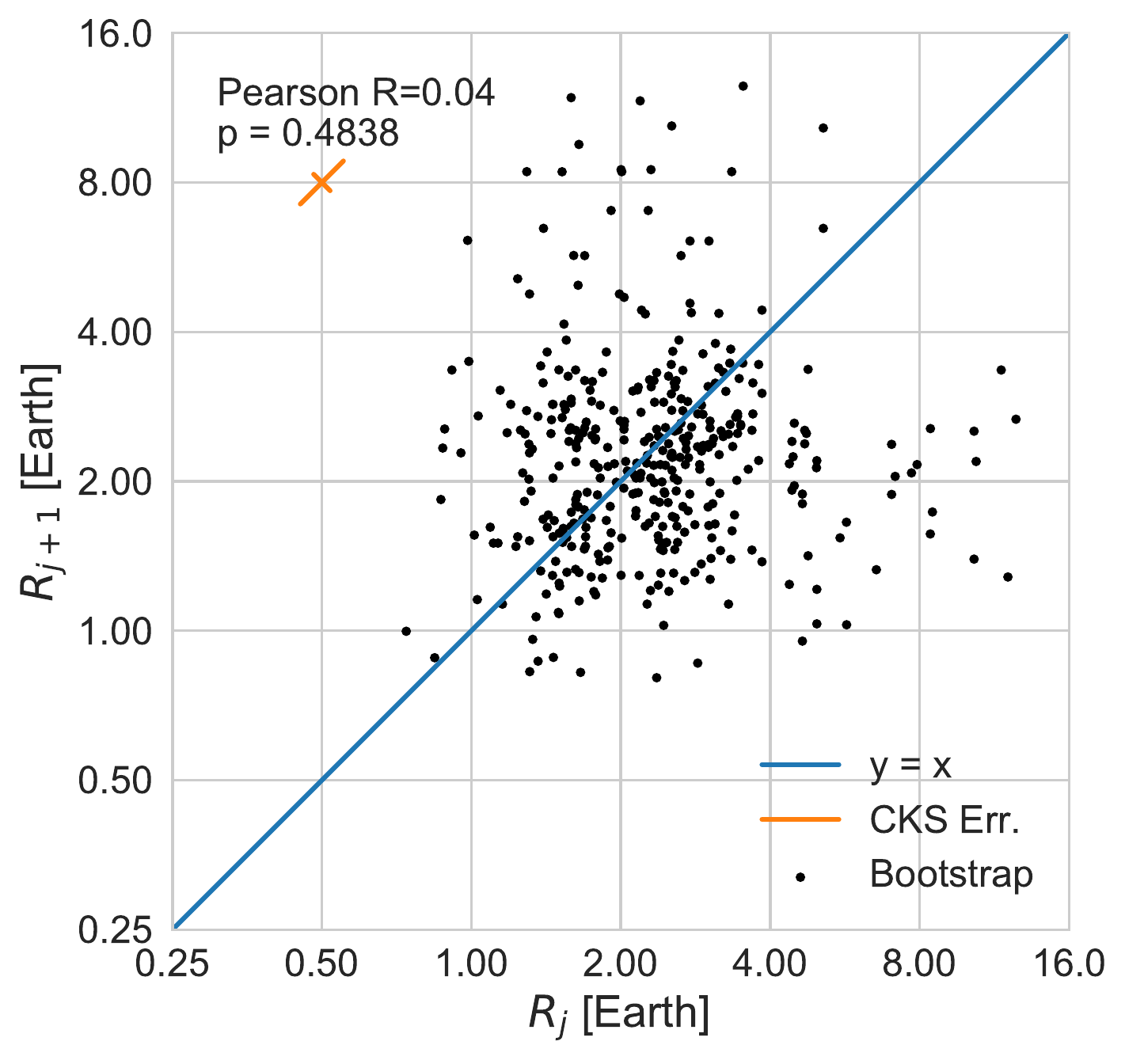} 
   \caption{The radius of a planet, $R_j$, vs. the radius of the next planet out, $R_{j+1}$, in one of one hundred bootstrap trials of the CKS multis (see text).  The bootstrap-generated distributions do not resemble the correlation between the sizes of adjacent planets in the CKS multis in Figure \ref{fig:rpcorr}.}
   \label{fig:rp_bootstrap}
   \end{center}
\end{figure}

\begin{figure}[htbp] 
   \begin{center}
   \includegraphics[width=\columnwidth]{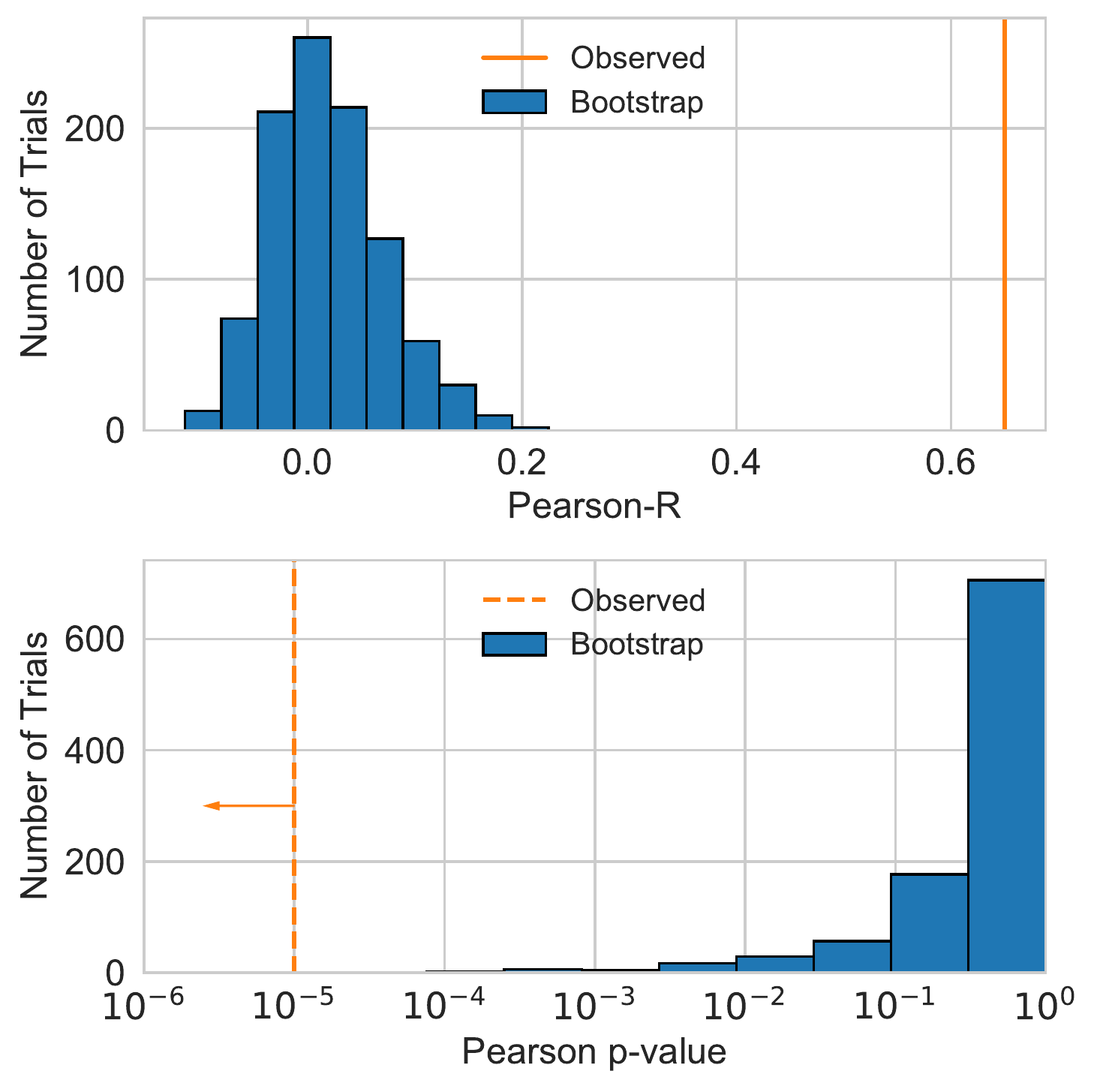}
   \caption{Left: The Pearson-R correlation between $R_j$ and $R_{j+1}$ in the observed distribution of CKS multis (orange line), as compared to the 1000 bootstrap trials drawn from the entire sample of 909 planets in the CKS multis (blue histogram).  Right: The p-value of the Pearson-R correlation between $R_j$ and $R_{j+1}$ in the observed distribution of CKS multis (orange dotted line, upper limit), as compared to the 1000 bootstrap trials (blue histogram).  The bootstrap trials do not exhibit the correlation between the inner and outer radii, demonstrating that the correlation is likely due to astrophysics, not detection biases.}
   \label{fig:rp_significance}
   \end{center}
\end{figure}

To further examine the role of detection biases in shaping the observed correlation, we conducted a series of bootstrap tests.  The null hypothesis underlying our bootstrap tests is that drawing each planet radius at random, with no regard for which star the planet originally orbited or which other planets were orbiting the star, and then subjecting this sample to the detection biases of \Kepler, can produce a correlation between $R_j$ and $R_{j+1}$.  The procedure was as follows:

\begin{enumerate}
\item Construct a bootstrap trial by drawing planet radii at random, \textit{with replacement}, from the distribution of observed planet radii\footnote{We also tried drawing radii from a lognormal distribution, with no appreciable difference in the results.} around all the CKS multis (see Figure \ref{fig:rp_dist}).  Note that this procedure has no regard for which star the planet originally orbited.  Note also that this procedure does not assume any relation between $R_j$ and $R_{j+1}$.
\item For each real CKS multiplanet host star, draw the number of planet radii equal to the number of transiting planets detected around that star, and place the planets at the observed orbital periods for that star.
\item Check that for each pair of planets, the smaller planet is detectable (SNR $>10$) when placed at the longer orbital period.  Discard any undetectable pairs.
\item Compute the Pearson-R correlation between the detectable pairs of $R_j$ and $R_{j+1}$. An example bootstrap result is shown in Figure \ref{fig:rp_bootstrap}.
\item Repeat 1000 times for each star, recording the Pearson-R value and p-value of each trial (Figure \ref{fig:rp_significance}).
\end{enumerate}

\begin{figure}[htbp]
	\includegraphics[width=\columnwidth]{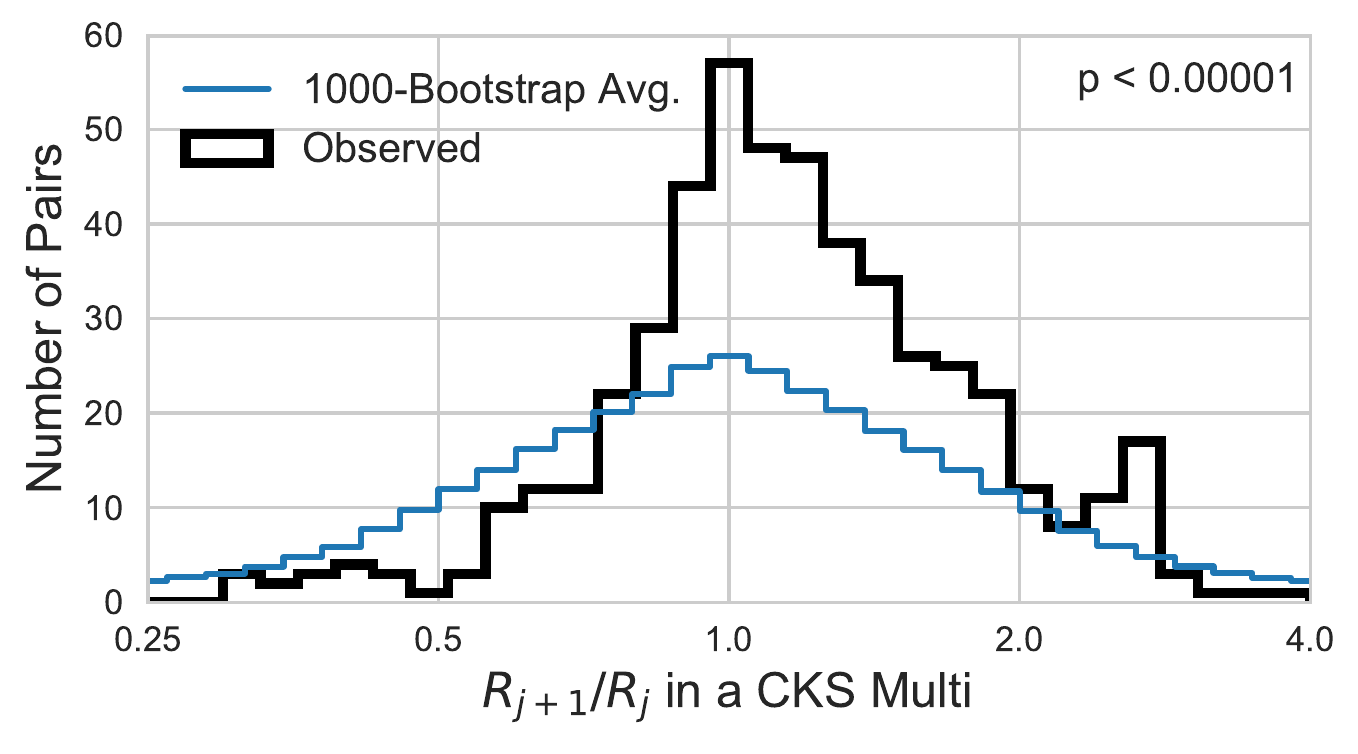}
\caption{The ratios of planet sizes for adjacent pairs within the same system (black line) compared to a control sample of detection-limited bootstrap tirals (blue line).  In both the observed and the bootstrap distributions, pairs are only counted if the smaller planet in the pair is detectable at the longer orbital period.  The $p$-value of an Anderson-Darling test comparing the observed vs. bootstrap distributions of planet radius ratios is $< 10^{-5}$; with a confidence of $>99.9999\%$, we can rule out the hypothesis that these two populations come from the same underlying distribution.  The distribution of observed planet radius ratios is significantly more peaked at 1.0 than the bootstrap distribution, indicating that planets in the same system are preferentially the same sizes.}
\label{fig:planet_sizes}
\end{figure}

Figure \ref{fig:rp_bootstrap} shows one example of a bootstrap trial and the resulting correlation.  The low Pearson-R value and near-unity p-value demonstrate a lack of correlation between adjacent planet sizes in this trial.  These values are typical of our 1000 trials, which are summarized in Figure \ref{fig:rp_significance}.  Our bootstrap tests were unable to reproduce the correlation between $R_j$ and $R_{j+1}$ pairs observed in real multi-planet systems.  

For example, the upper right corner of Figures \ref{fig:rpcorr} and \ref{fig:rp_bootstrap}---pairs of large planets, which are readily detectable---differ.  In the observed distribution, there are many pairs of large planets near (8, 8).  However, the bootstrap rarely realizes such pairs.  This is because only 7.8\% of the planets have $\rpl > 4~\rearth$, and so drawing two such planets from the distribution in a row is unlikely ($0.078^2 = 0.006$).  In the hypothesis underlying the bootstrap trials (each planet does not know about its neighbor), it is much more likely to draw a pair of planets with sizes of (2,8) or (8,2), since planets of 2~\rearth\ are very common.  

The discrepancy between the observed and bootstrap-generated planet pairs indicates that a null hypothesis influenced by detection biases cannot generate the observed correlation.  \textit{Therefore, the correlation between the sizes of adjacent planets is likely driven by astrophysics.}

To quantitatively compare the real and bootstrap distributions, we collapsed the 2D distributions into one dimension by computing $R_{j+1}/R_j$ for each pair.  Figure \ref{fig:planet_sizes} compares the observed and bootstrap-constructed distributions for the pairs of planet radii.  Using an Anderson-Darling test, we find a p-value of $< 10^{-5}$, allowing us to conclude with $>99.9999\%$ confidence that the observed pairs of planet radii are not drawn from the same distribution as the bootstrap-generated pairs of planet radii.   The observed distribution has a much sharper peak at a planet size ratio of 1.0 than the bootstrap distribution, meaning that planets in multi-planet systems are more likely to be similarly sized to each other than what we would expect drawing their sizes at random.  In the observed CKS multis, the distribution of $R_{j+1}/R_j$ has a mean value of \rpratmean, a median of \rpratmedian, and a standard deviation of \rpratstd.

The work presented here builds on previous studies of \Kepler\ multi-planet systems.  \citet{Lissauer2011_multis} also found that planets in multi-planet systems tend to be the same size, but with a much smaller sample of planets in multi-planet systems, resulting in only 71 independent, detectability-corrected ratios.  Our work repeats their experiment but with 504 independent, detectability-corrected ratios.  


\section{Planets in the same system have similar spacings}
\label{sec:spacing}
In Figure \ref{fig:architectures}, the planets appear evenly spaced in log semi-major axis.  Since log(A) - log(B) = log(A/B), constant differential log-spacing corresponds to a constant spacing ratio.

\begin{figure}[htbp] 
   \begin{center}
   \includegraphics[width=\columnwidth]{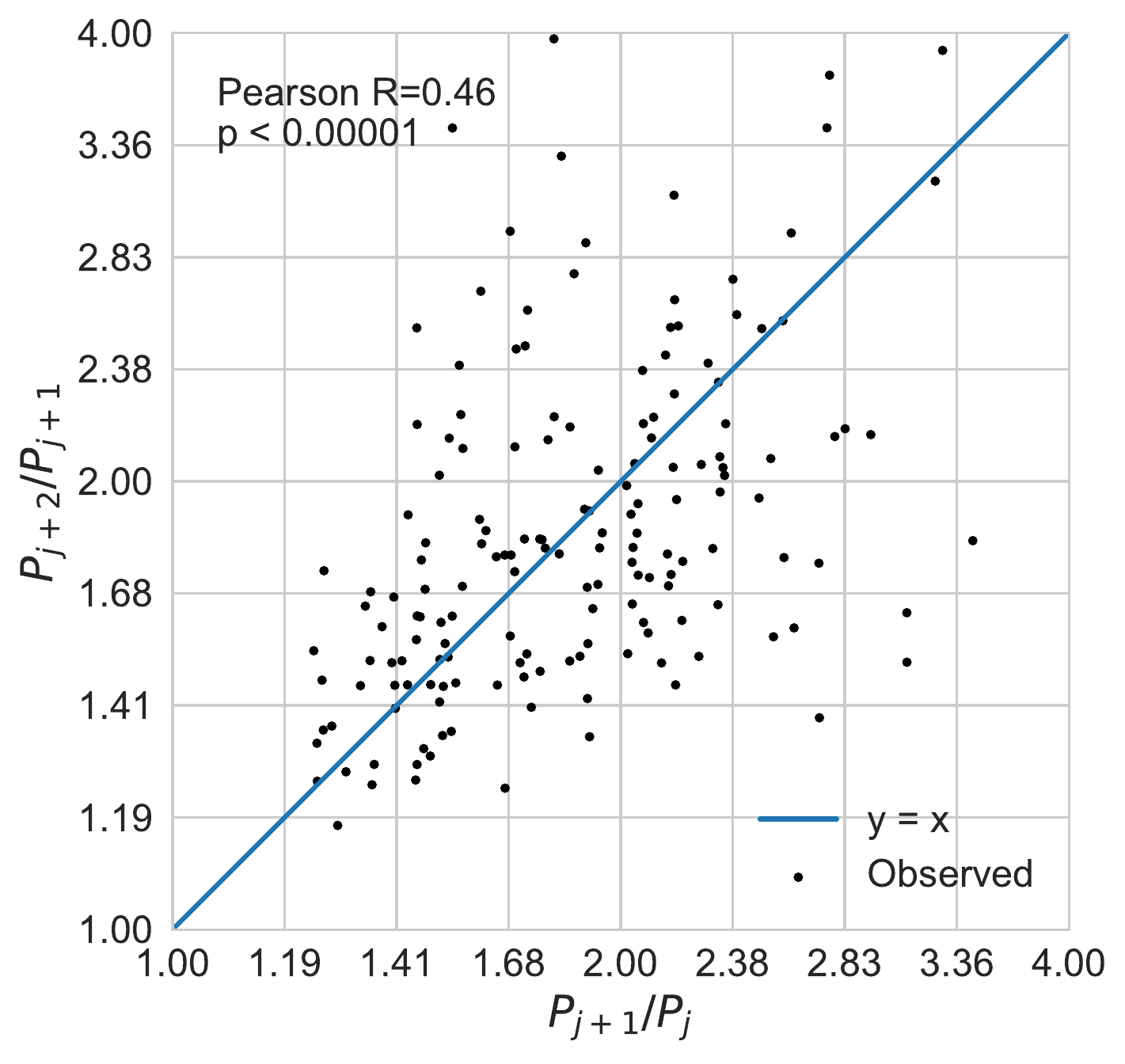} 
   \caption{In systems with three or more planets, the orbital period ratio of the outer periods vs. the orbital period ratio of the inner planets.  There is a large and statistically significant correlation (Pearson-R=0.46, $p < 10^{-5}$) between the orbital period ratios of the inner and outer planets.}
   \label{fig:prat_corr}
   \end{center}
\end{figure}

Within each observed system of three or more planets, we tested whether the orbital period ratio of adjacent planets, $P_{j+1}/P_j$, was correlated with the orbital period ratio of the next pair of planets out, $P_{j+2}/P_{j+1}$.  Since \Kepler\ observed for a finite amount of time, finding planets out to about 1000 days at most, our sensitivity to large ratios of orbital periods is incomplete.  If we find an example of $P_{j+1}/P_j = 10$, it is extremely unlikely that we would also be able to detect $P_{j+2}/P_{j+1} = 10$ in the same system, although a third planet might be detectable at $P_{j+2}/P_{j+1} = 2$.  Therefore, we limited our study to planet pairs with period ratios smaller than 4 (i.e. somewhat compact planetary systems; this number is justified in Section \ref{sec:size-spacing}).  These selection criteria resulted in 373 planets (165 pairs) around 104 stars.  The resulting orbital period ratios are shown in Figure \ref{fig:prat_corr}.  

In each group of three planets in a row, there is an apparent correlation between the orbital period ratio of the inner and outer pairs of planets.  We computed a Pearson-R correlation of 0.46, with a significance of p $< 10^{-5}$, which indicates that there is a strong, significant correlation between the orbital period ratios of planets in the same system.

\begin{figure}[htbp] 
   \begin{center}
\includegraphics[width=\columnwidth]{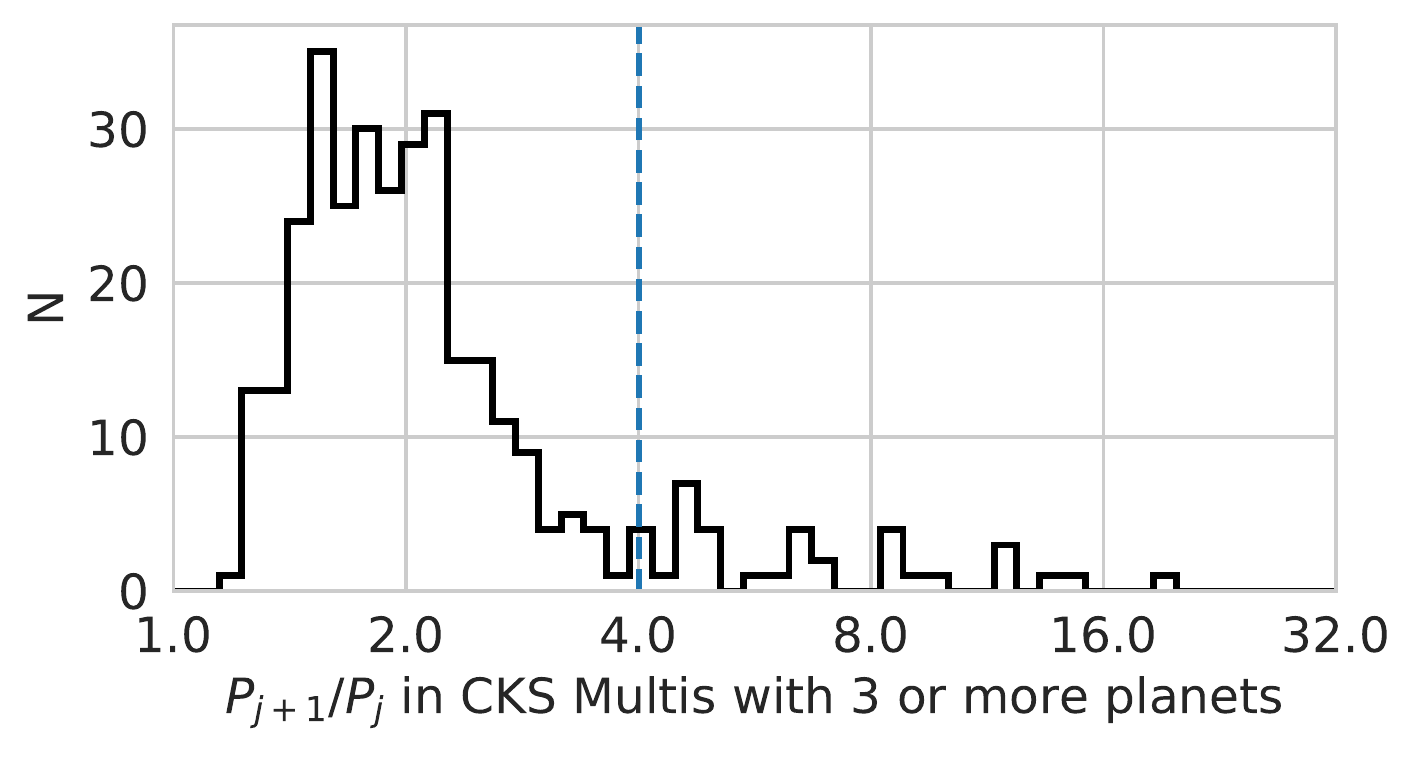}
   \caption{The distribution of the orbital period ratios of adjacent planets in the CKS multis.  Since \Kepler\ observed for a finite amount of time, we are more sensitive to patterns in which the orbital period ratios of the planets are small.  (A large orbital period ratio necessitates at least one long orbital period.)  Therefore, we limited our study to orbital period ratios smaller than 4 (dotted vertical line).}
   \label{fig:prat_dist}
   \end{center}
\end{figure}

\begin{figure}[htbp] 
   \begin{center}
   \includegraphics[width=\columnwidth]{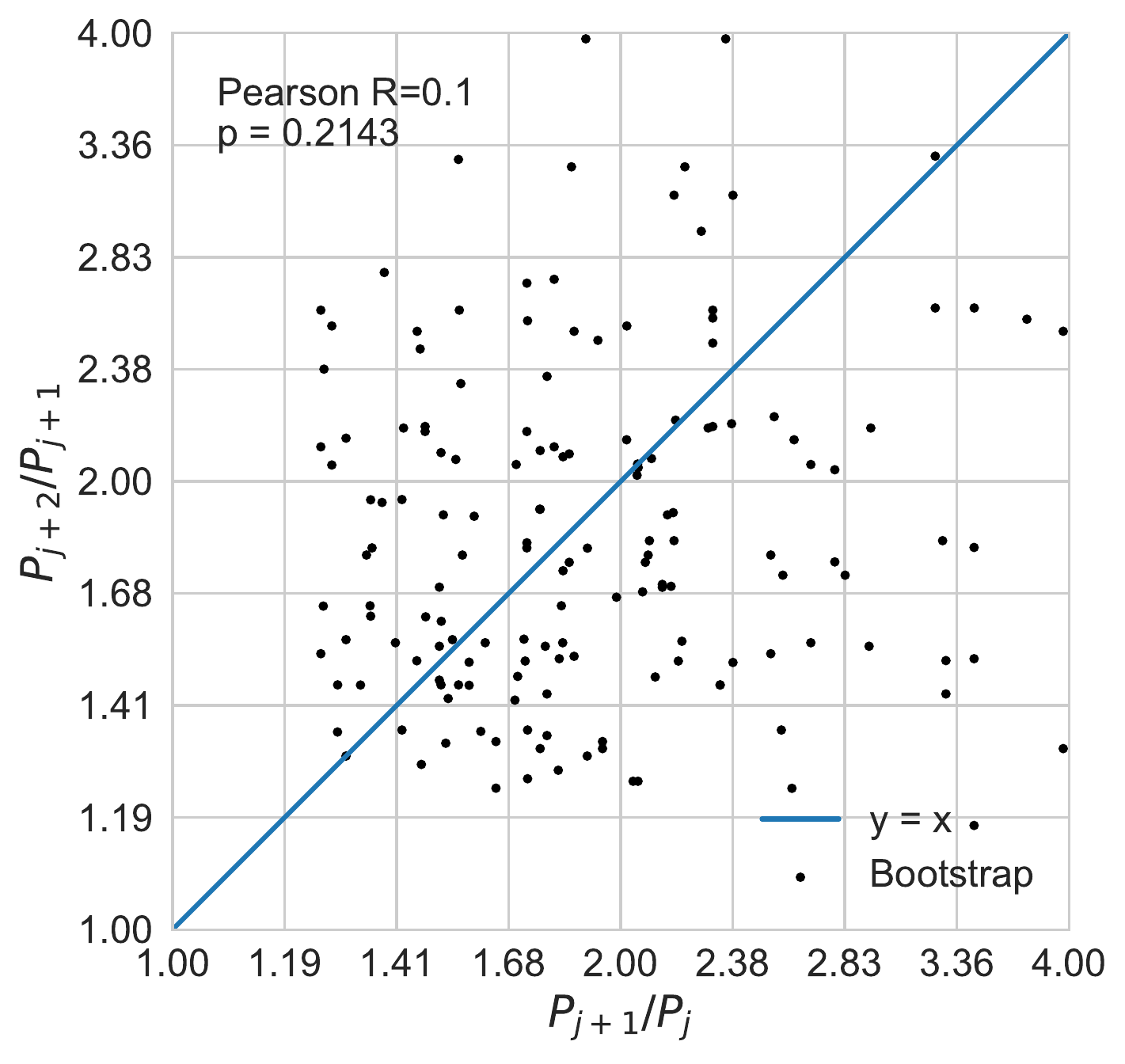} 
   \caption{The orbital period ratio of adjacent planets vs. the orbital period ratio of the next pair of adjacent planets in one of 1000 bootstrap trials of the CKS multis with 3 or more transiting planets (see text).  The lack of correlation between the spacings of adjacent planet pairs in the bootstrap trial is different from the strong correlation in the distribution of observed orbital period ratios (Figure \ref{fig:prat_corr}).}
   \label{fig:prat_bootstrap}
   \end{center}
\end{figure}

\begin{figure}[htbp] 
   \centering
   \includegraphics[width=\columnwidth]{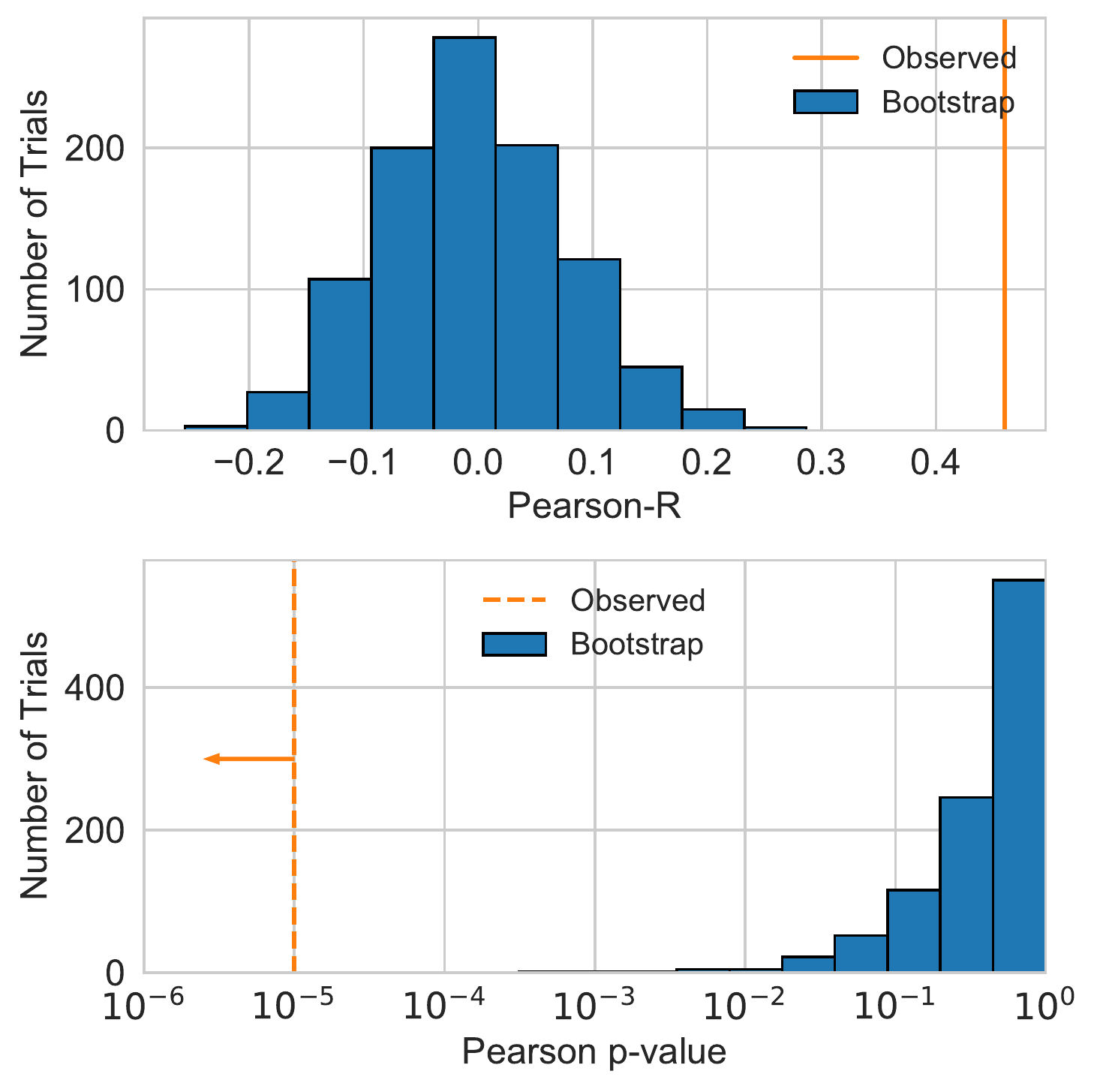}
   \caption{Left: The Pearson-R correlation between the inner ($P_{j+1}/P_j$) and outer ($P_{j+2}/P_{j+1}$) period ratios in the observed distribution of CKS multis with 3 or more planets (orange line), as compared to the 1000 bootstrap trials (blue histogram).  Right: The p-value of the Pearson-R correlation between the period ratios in the observed distribution of CKS multis (orange dotted line, upper limit), as compared to the 1000 bootstrap trials (blue histogram).  The bootstrap trials do not reproduce the significance of the observed correlation, demonstrating that the correlation is likely due to astrophysics, not detection biases.}
   \label{fig:prat_significance}
\end{figure}

As above, we used a bootstrap analysis to test whether the observed correlation in adjacent orbital periods could arise from detection biases.  The null hypothesis we formulated was that the orbital period ratio between one pair of planets was not, a priori, related to the orbital period ratio of the next pair of planets in the same system.  Our procedure for constructing the bootstrap trials was as follows:
\begin{enumerate}
\item Construct a bootstrap trial by drawing each orbital period ratio at random, \textit{with replacement}, from the distribution of observed orbital period ratios smaller than 4 (Figure \ref{fig:prat_dist}).  Note that this procedure has no regard, a priori, for which star the planet pair originally orbited, or what the orbital period ratios among other planets in the system might be.
\item For each real CKS multiplanet host star, draw the number of orbital period ratios equal to the number of transiting planets observed around that star, minus one.
\item To map the orbital period ratios to orbital periods, draw an orbital period $P_1$ at random, \textit{with replacement}, from the observed distribution of orbital periods of the innermost transiting planet in each system.
\item Multiply the orbital period $P_1$ by the first drawn orbital period ratio to get the second orbital period, $P_2$.  Repeat (replacing $P_1$ with $P_2$, etc.) until each planet has been assigned an orbital period.
\item Remove any planets with orbital periods greater than 1071 days (the maximum observed orbital period of a planet in the CKS multis).
\item Remove any planets that are not detectable (i.e., SNR $\le10$) at their randomly assigned orbital periods.
\item Among the remaining (detectable) planets in the bootstrap trial, compute the orbital period ratios for each adjacent pair of planets.  
\item Compute the Pearson-R correlation between each inner ($P_{j+1}/P_j$) and outer ($P_{j+2}/P_{j+1}$) orbital period ratio, in systems that retain three or more detectable planets.  Figure \ref{fig:prat_bootstrap} shows one example.
\item Repeat 1000 times, recording the Pearson-R and p-value of each bootstrap trial (Figure \ref{fig:prat_significance}).
\end{enumerate}

Figure \ref{fig:prat_bootstrap} shows one example of a bootstrap trial and the resulting correlation.  The low Pearson-R value and near-unity p-value demonstrate a lack of correlation between the orbital period ratios of adjacent pairs of planets in this trial.  These values are typical of our 1000 trials, which are summarized in Figure \ref{fig:prat_significance}.  Our bootstrap tests were unable to reproduce the distribution of $P_{j+1}/P_j$ and $P_{j+2}/P_{j+1}$ pairs observed in real multiplanet systems, indicating that detection biases alone cannot generate the observed correlation.  \textit{Therefore, the correlation between the spacings of adjacent pairs of planets is likely driven by astrophysics.}

\begin{figure}[htbp]
	\includegraphics[width=\columnwidth]{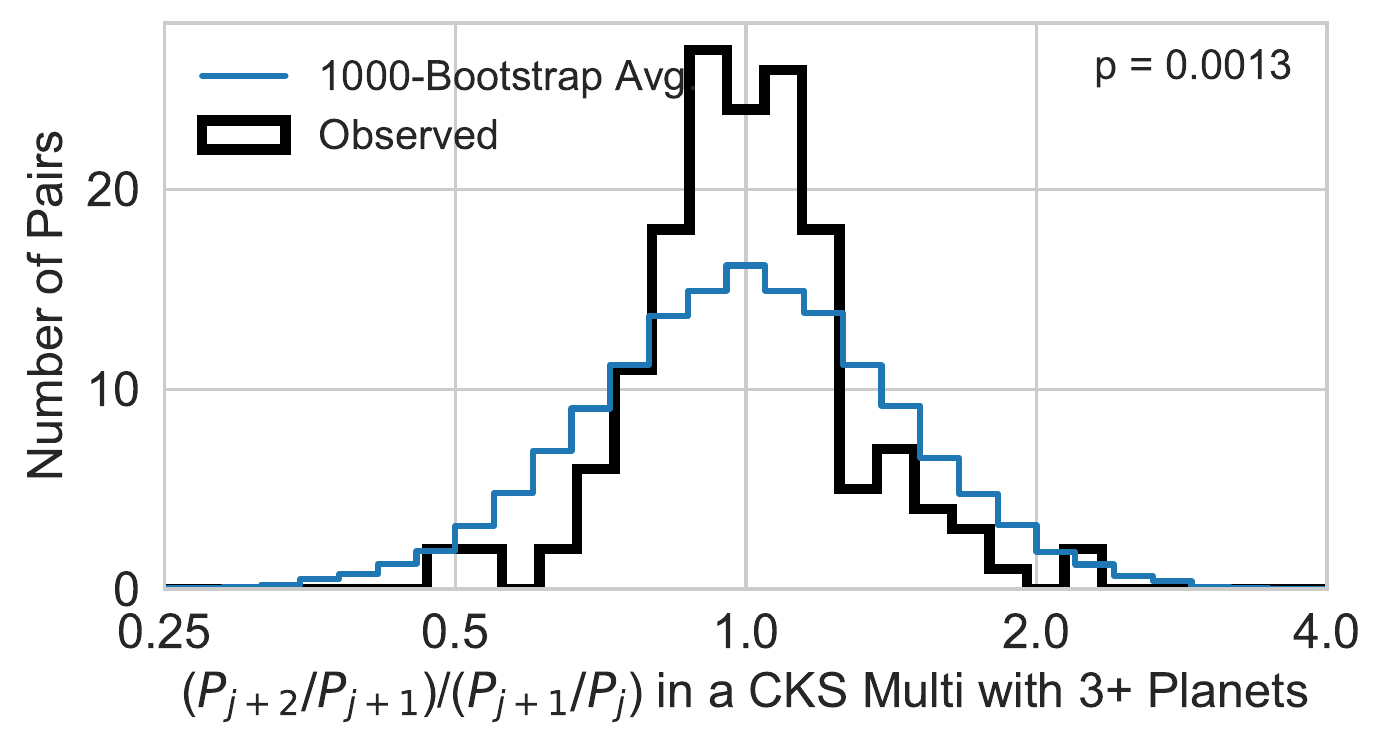}
\caption{The ratio of orbital period ratios for each adjacent triple of planets within the same multi-planet system (black line) compared to a control sample of detection-limited bootstrap trials (blue line).  In both the observed and the bootstrap distributions, pairs are only counted if the smaller planet in the pair is detectable at the longer orbital period.  The $p$-value of an Anderson-Darling test comparing the observed vs. bootstrap distributions of planet radius ratios is 0.0013; with a confidence of $>99\%$, we can rule out the hypothesis that these two populations come from the same underlying distribution.  The distribution of observed ratios of orbital period ratios is significantly more peaked at 1.0 than the bootstrap distribution, indicating that planets in the same system have a preferred spacing.}
\label{fig:planet_spacings}
\end{figure}

To quantitatively compare the real and bootstrap distributions, we collapsed the 2D distributions into one dimension by computing $\mathcal{P} = (P_{j+2}/P_{j+1})/(P_{j+1}/P_{j})$ for each pair.  Figure \ref{fig:planet_spacings} compares the observed and bootstrap-constructed distributions for the pairs of planet radii.  Using an Anderson-Darling test, we find a p-value of $0.0012$, allowing us to conclude with $99.9\%$ confidence that the observed pairs of planet radii are not drawn from the same distribution as the bootstrap-generated pairs of planet radii.   The observed distribution has a sharper peak at a ratio of ratios of orbital periods of 1.0 than the bootstrap distribution, meaning that planets in multiplanet systems are more likely to have correlated spacings than what we would expect drawing their orbital period ratios at random.  The lower significance of this correlation than for the correlation between adjacent planet sizes likely stems from the smaller number of pairs (165) than were available for the study of adjacent planet radii (504), since comparing orbital period ratios requires at least three planets in a system.  In the observed CKS systems of 3 or more planets, the distribution of $\mathcal{P}$ has a mean value of \pratratmean, a median of \pratratmedian, and a standard deviation of \pratratstd.

\citet{Malhotra2015} and \citet{Steffen2015} have also characterized the period ratio distributions of the \Kepler\ multis.  \citet{Malhotra2015} noted that the distribution of planet spacings, $\mathcal{D} = 2\frac{a_2 - a_1}{a_1 + a_2}$, is approximately log-normal.  Our result differs from theirs in that we explored how the orbital spacing between one pair of planets and the next pair in the same system are correlated.  \citet{Steffen2015} also examined the relationships between the period ratios of adjacent pairs of planets, but only for those in which one pair (or the product of the pairs) was near 2.2.
\section{Physical Underpinnings}
\label{sec:underpinnings}
How do planets know to be the same size?  How do they know how far apart to form?  Below, we explore the relationship between planet size and spacing and estimate the planet spacings in terms of mutual Hill radii. We also examine how stellar incident flux relates to planet size ratios.  

\subsection{The Relation between Planet Size and Spacing}
\label{sec:size-spacing}
\begin{figure}[htbp] 
   \centering
\includegraphics[width=\columnwidth]{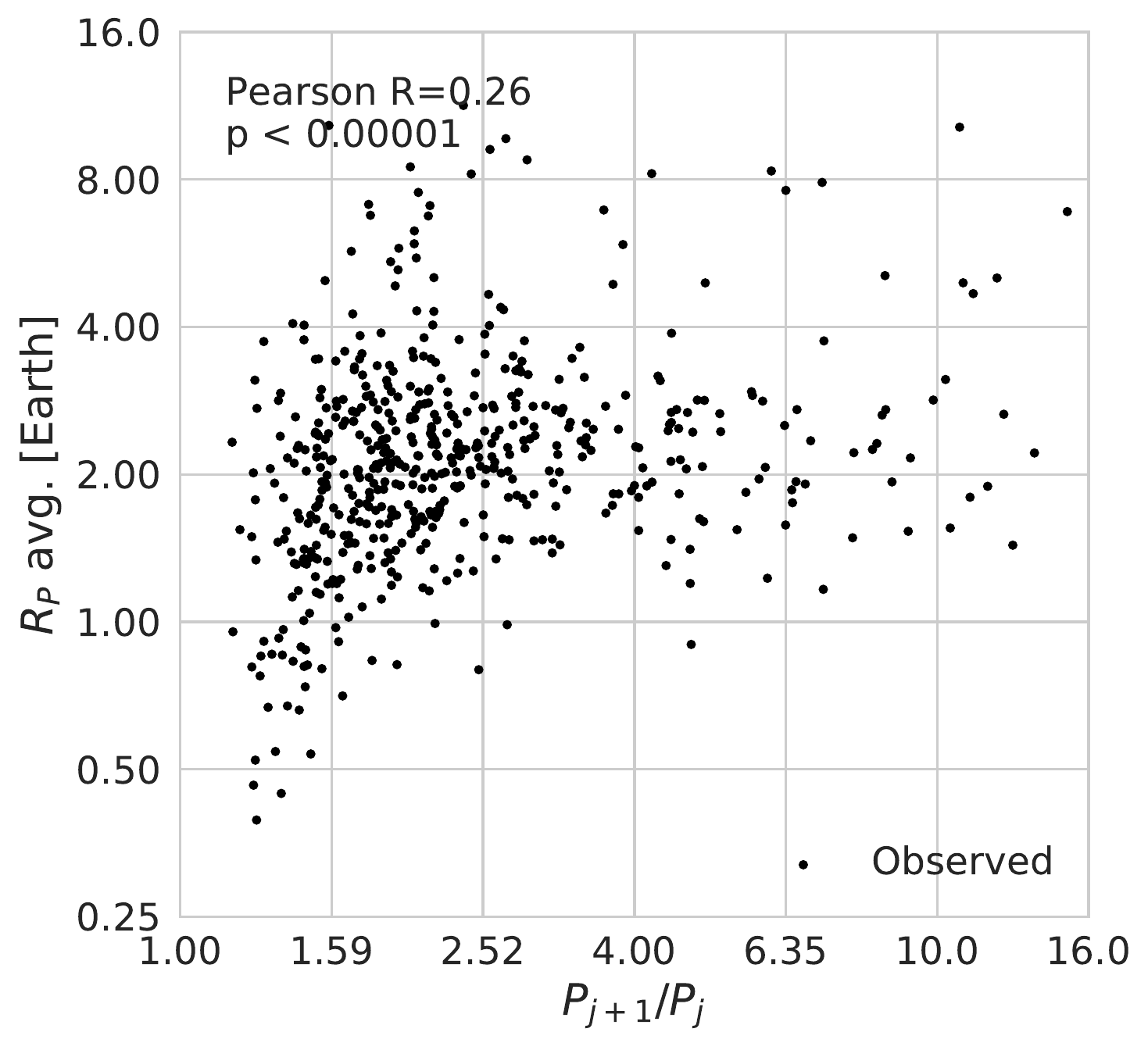}
   \caption{The average planet size in an adjacent pair ($R_P$ avg) vs. the orbital period ratio of the pair, for the CKS multis.  There is a slight positive correlation (Pearson-R = 0.26, p-value $< 10^{-5}$) between the average planet size and the orbital period ratio.  The relationship is strongest at small period ratios.}
   \label{fig:size-spacing}
\end{figure}

In Figure \ref{fig:architectures}, the systems with the smallest planets appear to have the closest spacings, and the systems with larger planets appear to have larger spacings.  Extending this observation to the full sample of CKS multis, we find a correlation between the sizes of planets and their orbital period ratios (Figure \ref{fig:size-spacing}).  There is a statistically significant positive correlation ($r=0.26$, $p<10^{-5}$) between the orbital period ratio and average planet size in a pair.  The correlation is particularly visible for $P_{j+1}/P_j < 4$, over which range the planet size increases with increasing orbital period ratio.

Two features in Figure \ref{fig:size-spacing} draw the eye.  There is a striking wall at $P_{j+1}/P_j = 1.2$: essentially no planets are closer together than this.  The wall spans an order of magnitude in planet size, extending from the smallest planets ($\rpl \approx 0.5~\rearth$) to Neptune-sized planets.  Also, there is an absence of small planets at large orbital period ratios.

We investigated whether some artifact of the \Kepler\ pipeline for multi-planet systems could produce the wall at $P_{j+1}/P_j = 1.2$.  According to \citet{Tenenbaum2013}, the procedure for investigating a light curve in which one transiting planet candidate has already been found is as follows: ``The transit signatures from the fitted planet model are removed from the flux time series; and the residual flux time series is then searched for additional TCEs.''  There is no mention of any period filtering that would exclude planets that are closer together than a period ratio of 1.2.  Furthermore, planets that have a period ratio of 1.2 (or even 1.1) but are initially overlapping in transits quickly become out of phase, and so after a few orbits the transits occur in different places in the light curve.  Hence, we do not identify any aspect of the \Kepler\ pipeline that would produce the wall at a period ratio of 1.2.

The period ratio boundary at 1.2 was predicted by \citet{Deck2013}, who used numerical integrations and analytical calculations to demonstrate that when planets get closer together than $\frac{a_2 - a_1}{a1} < 1.46 (\mu_1 + \mu_2)^{2/7}$ (where $\mu$ is the planet-to-star mass ratio), they are very likely to stumble into a resonance overlap of two or more first-order mean motion resonances, leading to dynamical chaos and eventual Lagrange instability.  Even though low-mass planets might be Hill stable in these configurations, their entanglement in resonance overlap is fatal for stability.  For planets of $\mpl/\mstar \approx 10^{-5}$, i.e., roughly 3~\mearth, with low eccentricities ($e < 0.04$), the typical minimum Lagrange-stable separation corresponds to a period ratio of about 1.2 \citep[][Figure 12]{Deck2013}.  Although some planets can survive closer than this separation if they fall in an island of stability\footnote{based on serendipitous initial relative mean anomalies}, there are very few planets for which this appears to be relevant\footnote{Kepler-36 is one such exception and was scrutinized in \citet{Deck2013}}.

We investigated the absence of small planets at large orbital period ratios.  The smallest planets (\rpl $< 1~\rearth$) are the closest to each other, with typical orbital period ratios from 1.2 to 2.0.  These sub-Earths orbit 22 stars, and so the clustering of all these planets at small orbital period ratios is not a coincidence derived from a small number of systems with compact architectures.  Among the sub-Earths alone (45 pairs), the Pearson-R coefficient is 0.44, with $p=0.002$.

\begin{figure}[htbp] 
   \centering
\includegraphics[width=\columnwidth]{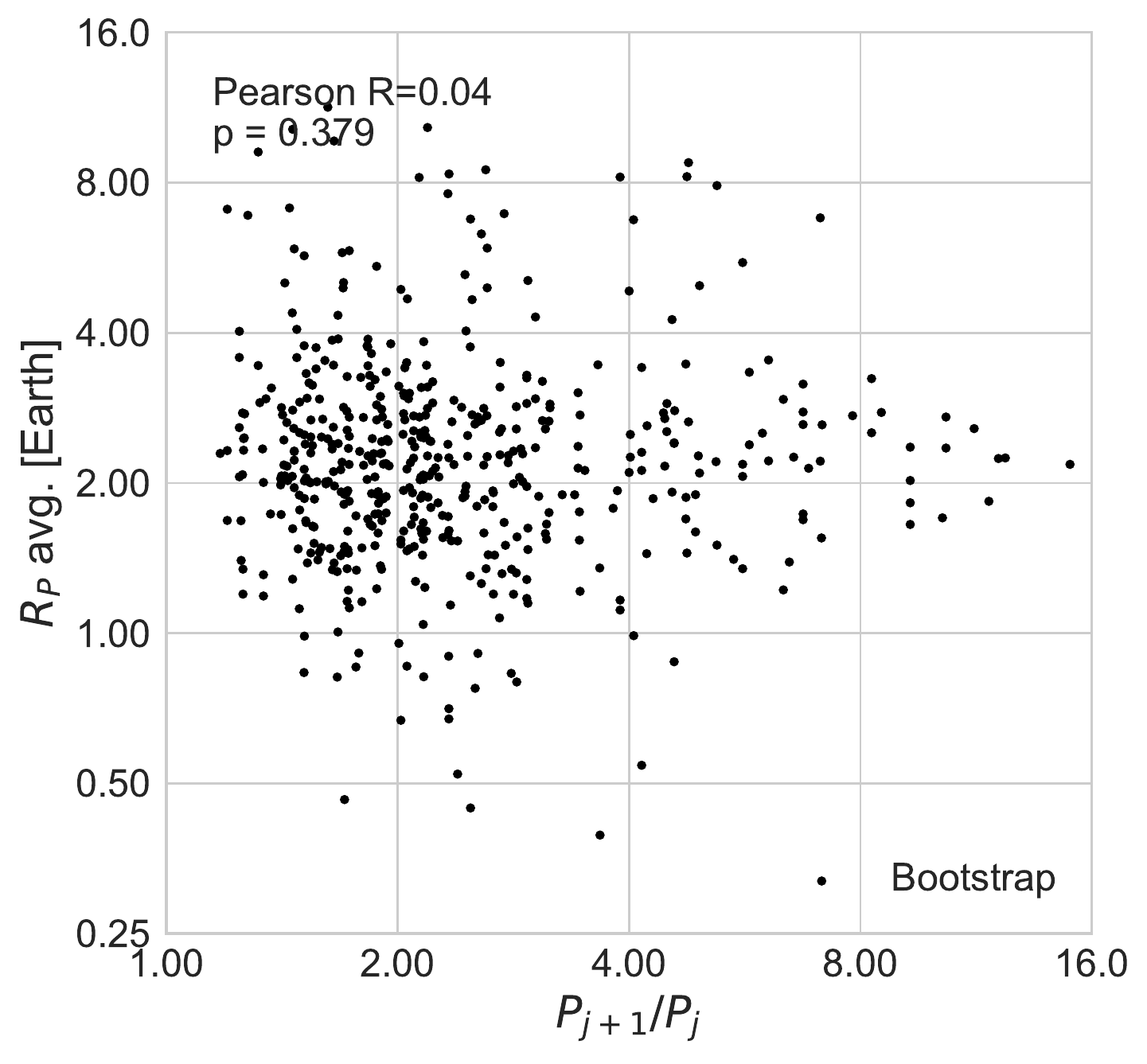}
   \caption{The average planet size in an adjacent pair ($R_P$ avg) vs. the orbital period ratio of the pair in one bootstrap realization.  There is no correlation between the average planet size and the orbital period ratio.  The smallest planets ($\rpl < 1~\rearth$) are detected at orbital period ratios of 4, in this example, demonstrating that the structure in the lower left of Figure \ref{fig:size-spacing} is not sculpted by detection biases.}
   \label{fig:size-spacing-bootstrap}
\end{figure}

Is the correlation between orbital period ratio and planet size astrophysical?  We used the bootstrap method of Section \ref{sec:spacing} to test how moving the small planets to a variety of orbital separations would affect their detectability.  Although planets smaller than 1~\rearth\ are detectable around their parent stars at $P_{j+1}/P_j \sim 4$ (Figure \ref{fig:size-spacing-bootstrap}), no such planets have been found.  None of 1000 bootstrap trials generated a correlation with larger Pearson-R or smaller p-value than the observed distribution.  Thus, the lack of small planets at orbital period ratios of $\sim4$ in our sample is probably not solely based on detection bias, since we could have found such planets.  Therefore, the clustering of small planets at very close orbital period ratios in Figure \ref{fig:size-spacing} is very likely sculpted by astrophysics.

\subsection{Planets are 20 Mutual Hill Radii Apart}
One astrophysical interpretation that links planet sizes and spacing is the idea that planets communicate with their neighbors through gravitational interactions.  We therefore considered the separation between a pair of planets in terms of their gravitational influence, of which the mutual Hill radius is the natural unit.  The mutual Hill radius of two planets of masses $m_j$ and $m_{j+1}$ orbiting a star of mass $\mstar$ at semi-major axes $a_j$ and $a_{j+1}$ is
\begin{equation}
R_H = \Big(\frac{m_j + m_{j+1}}{3\mstar}\Big)^{1/3} \frac{(a_j + a_{j+1})}{2}
\label{eqn:mrh}
\end{equation}
and the separation between two planets, in units of mutual Hill radii, is
\begin{equation}
\Delta = (a_{j+1} - a_j)/R_H
\label{eqn:drh}
\end{equation}
\citep{Gladman1993}.  

It is useful to note that $\Delta$ is related to but not directly proportional to the orbital period ratio, $\mathcal{P} \equiv P_{j+1}/P_j$.  Using Kepler's third law to rewrite equations \ref{eqn:mrh} and \ref{eqn:drh} in terms of orbital periods, we have

\begin{equation}
\Delta = 2 \Big(\frac{m_j + m_{j+1}}{3\mstar}\Big)^{-1/3} \Big(\frac{\mathcal{P}^{2/3} - 1}{\mathcal{P}^{2/3} + 1}\Big)
\end{equation}

Since the period ratios we are considering tend to be small ($\sim 1$ to 4), subtracting one in the numerator and adding one in the denominator significantly alters the fraction, making $\Delta$ distinct from the period ratio, especially for the smallest planets, which all have $P_{j+1}/P_j < 2$.

To compute planet separations in mutual Hill radii, it is necessary to adopt planet masses.  We converted our precise planet radii to estimates of planet masses ($\mpl$) and densities ($\rhopl$) via the empirical mass-radius relationships of \citet{Weiss2014} and \citet{Weiss2013}:
\renewcommand{\thefootnote}{\roman{footnote}}
\begin{align} \label{m-r}
\rpl/\rearth & < 1.5: \nonumber\\
                 & \rhopl = 2.43 + 3.39 (\rpl/\rearth) \gcc \nonumber\\
                  & \mpl = (\rhopl / 5.51 \gcc) (\rpl/\rearth)^3~\mearth
\end{align}
\begin{align}
1.5 & \le \rpl/\rearth \le 4.0:  \nonumber\\
      & \mpl = 2.69 (\rpl/\rearth)^{0.93}~\mearth
\end{align}
\begin{align}
4.0 & < \rpl/\rearth < 9.0:  \nonumber\\
      & \mpl = 0.86 (\rpl/\rearth)^{1.89}~\mearth\footnotemark 
\end{align}
\begin{align}
9.0 & \le \rpl/\rearth:  \nonumber\\
       & \mpl = 100~\mearth \text{; i.e., } \mjup\footnotemark
\end{align}
\footnotetext[1]{The original formulation of this relation includes a very weak dependence on the incident stellar flux ($(F/\fearth)^{0.057}$). Since this weak dependence might not be valid, we apply $F=100~\fearth$ as a substitute.} 
\footnotetext[2]{The masses of Jupiter-sized planets vary widely.  Because only 2\% of the planets are in this size range, the results are insensitive to the masses we assume for these planets.}

The calculated mutual Hill radii, along with other useful system properties, are available in Table \ref{tab:multis}.  Although there is large scatter in the planet masses with respect to these mean empirical relations \citep{Marcy2014, Weiss2014, Rogers2015, Wolfgang2015, Chen2017}, we have no reliable basis for deciding which planet masses should be higher or lower than the mean, and so we adopt a simple one-to-one mapping of radius to mass.  Furthermore, since the Hill radius scales as $m^{1/3}$, uncertainties in the mass of order a few do not seriously affect the estimated mutual Hill radius.

\begin{figure}[htbp]
\begin{center}
\includegraphics[width=\columnwidth]{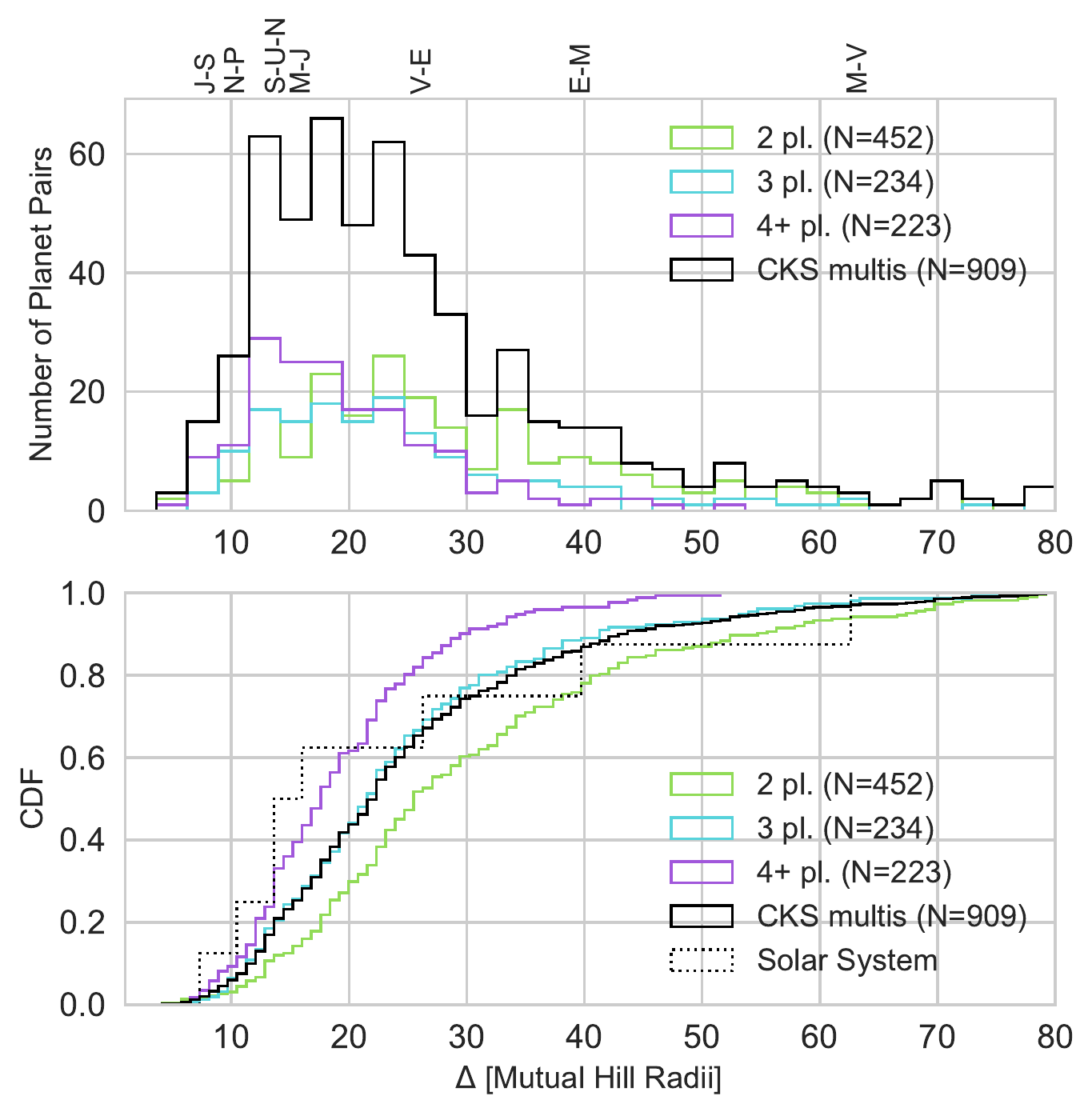}
\caption{Top: Separation in mutual Hill radii between adjacent pairs of transiting planets, assuming the empirical mass-radius relations from \citet{Weiss2014} and \citet{Weiss2013}.  The CKS multis are shown (black line), as are the sub-samples with two (green), three (cyan), or at least four (purple) transiting planets.  Planet pairs in the solar system are shown as dotted lines, with the planet names at the top of the plot.  Bottom: same as top, but showing the cumulative distribution function.}
\label{fig:hill_sep}
\end{center}
\end{figure}

The orbital separations in mutual Hill radii are shown in Figure \ref{fig:hill_sep}.  The top panel is a histogram of the CKS multis, including sub-samples of the CKS multis with two, three, or at least four transiting planets, and annotations corresponding to the mutual Hill radii of solar system planets.  The bottom panel is the cumulative distribution function.  The majority of \Kepler\ planets (93\%) are at least 10 mutual Hill radii apart, and the distribution of mutual Hill radii peaks at around 20.  \citet{Fang2012}, \citet{Pu&Wu2015}, and \citet{Dawson2016} have also estimated that the Kepler planets have a typical spacing of about 20 mutual Hill radii, \textit{but our updated planet radii and stellar masses allow more precise empirical estimates.}

Systems with high multiplicity of transiting planets (4+) tend to have the smallest Hill separations, shown in purple.  For example, the Kepler-11 planets have $\Delta < 10$.  However, the masses of these planets are known: the Kepler-11 planets all have systematically low densities, making their masses smaller than what we predicted from a simple mass-radius relationship.  Because many compact systems with TTVs have systematically low densities \citep{Weiss2014}, a more sophisticated mass-radius relation is necessary to quantitatively link the compactness of a system, the planet radii, and the planet masses.

For systems that have larger Hill separations, it is possible that the eccentricities and/or mutual inclinations of the planets are larger, requiring larger separations between the planets for stability.  For systems with pairs of planets more than about 20 mutual Hill radii apart, it is also possible that another planet resides between them, but either does not transit or has not been detected.  It also possible that systems with large dynamical separations are not maximally packed.

\begin{figure}[htbp] 
   \centering
   \includegraphics[width=\columnwidth]{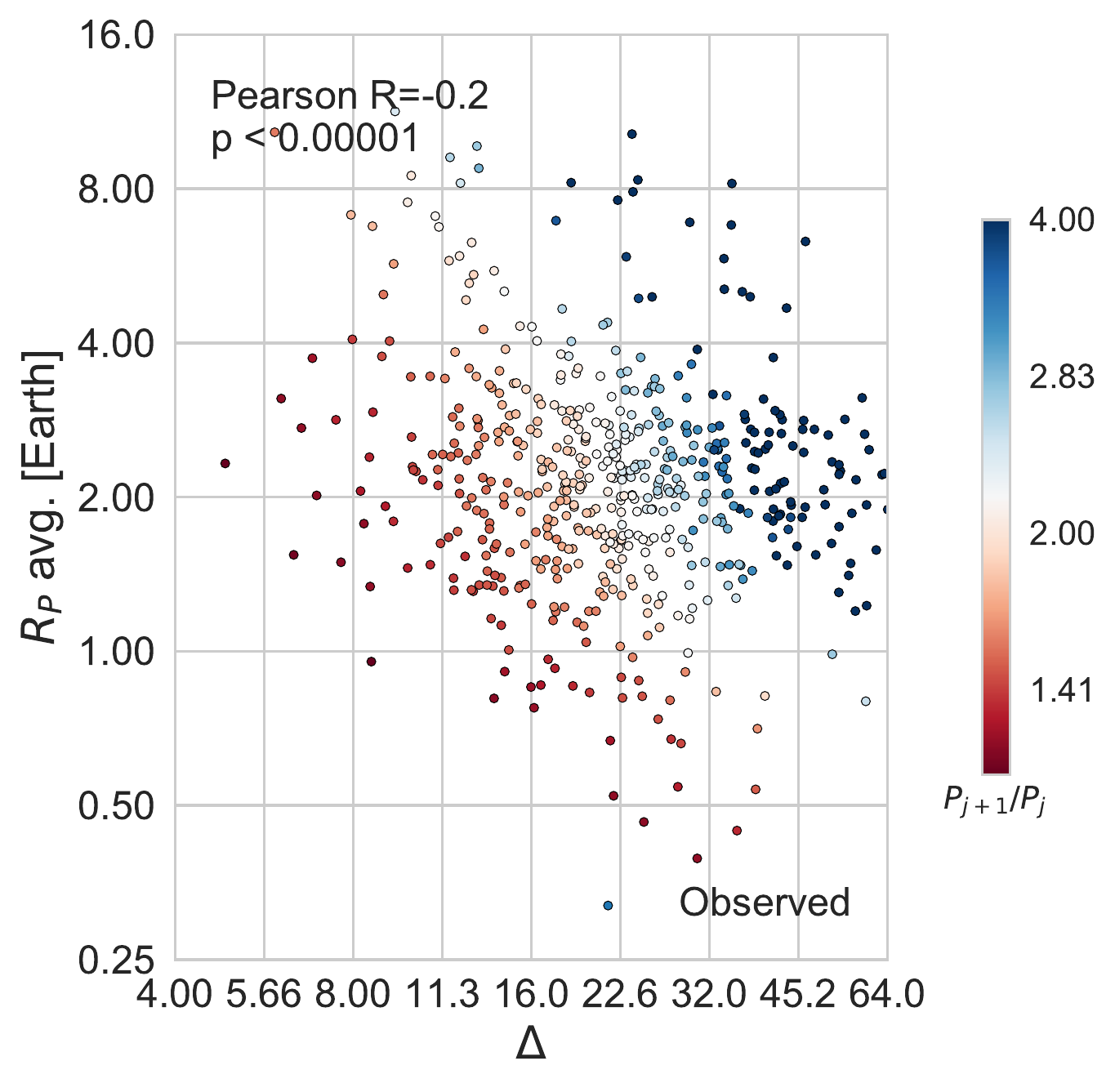} 
   \caption{The average planet radius in an adjacent pair, $R_P$ avg., vs. the estimated spacing in mutual Hill radii between those planets, $\Delta_\mathrm{RH}.$  The colors correspond to the orbital period ratio, $P_{j+1}/P_j$.  The negative correlation between planet size and mutual Hill radius ($r = 0.26$, $p < 10^{-5}$) appears to arise from an absence of points in the lower left corner of the plot.  Specifically, planets smaller than $1~\rearth$ are at least 16 mutual Hill radii apart.  The absence in the lower left corner is related to the absence of planets closer together than a period ratio of 1.2.}
   \label{fig:rp-hill-color}
\end{figure}

If the mutual Hill radius, rather than the orbital period ratio, is the fundamental unit underlying planet spacing, we should expect no correlation between separation in mutual Hill radii and planet size.  The mutual Hill radius incorporates the mass of the planet, and so ideally, the computation of mutual Hill radius separation should remove the contribution of planet size.  Figure \ref{fig:rp-hill-color} shows the separation in mutual Hill radii vs. the average planet size.  A Pearson-R test finds a significant negative correlation between planet size and separation in mutual Hill radii ($r = -0.2$, $p < 10^{-5}$).  This correlation is driven by an absence of points in the lower left corner of the plot.  Planets smaller than $1~\rearth$ are all farther apart than 16 mutual Hill radii.  It would have been easier to detect these planets if they were closer together (i.e., if they had shorter orbital periods), and so the absence of very small planets at close dynamical spacings is not due to detection bias.

The colors in Figure \ref{fig:rp-hill-color} indicate the orbital period ratio of each pair.  The closest pairs, with $P_{j+1}/P_j\approx 1.2$, are red.  The colors run parallel to the lower left edge, suggesting that the absence of points in the lower left corner is related to the absence of any planets closer together than an orbital period ratio of 1.2 (Figure \ref{fig:size-spacing}).

\subsection{Temperature difference correlates with planet size ratios}
In $65.4\pm0.4\%$ of planet pairs in the CKS multis, the outer planet is larger than the inner planet.  We investigate whether incident stellar flux is correlated with the asymmetry of planet sizes.  For each adjacent pair of planets, we consider the difference in their estimated equilibrium temperatures based on the incident stellar flux at each planet\footnote{assuming a bond albedo of 0.3, as in CKS II}: $\Delta T_\mathrm{eq} = T_\mathrm{inner} - T_\mathrm{outer}$.  

\begin{figure}[htbp] 
   \centering
\includegraphics[width=\columnwidth]{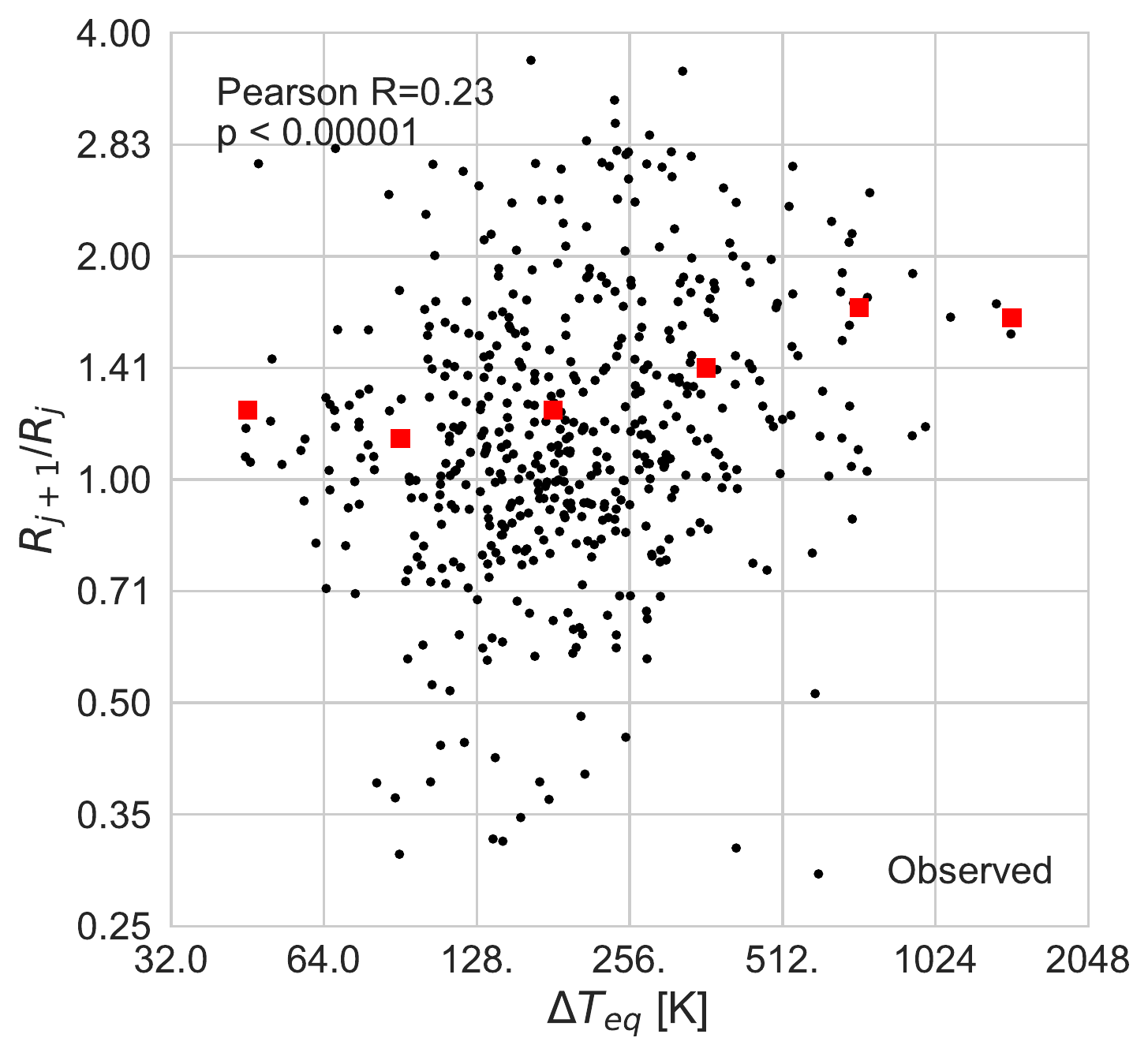}
   \caption{The ratio of adjacent planet sizes vs. their difference in expected equilibrium temperatures.  There is a slight positive correlation (Pearson-R = 0.23, p-value $< 10^{-5}$) between the temperature difference and the planet size ratio.  To guide the eye, averages in bins of log(d$T$) are shown as red squares.  This correlation is consistent with photo-evaporation, which should result in inner planets being smaller than outer planets.}
   \label{fig:teq-corr}
\end{figure}

In Figure \ref{fig:teq-corr}, we investigate whether the difference in equilibrium temperature is correlated with the size ratio of the outer to inner planet.  (The size ratio is simply $y/x$ from Figure \ref{fig:rpcorr}.)  We find a slight correlation: the greater the temperature difference between the inner and outer planet, the larger the ratio of their sizes.  In particular, there is an absence of planet pairs with a larger inner planet when the temperature difference is large.  Photo-evaporation \citep[e.g.,][]{Lopez2012,Owen2013,Zahnle2017} could produce such an effect, since the expectation from photo-evaporation is that inner planets should be smaller than outer planets (assuming identical core masses), and that this effect should be more pronounced when the temperature difference is larger.

\begin{figure}[htbp] 
   \centering
\includegraphics[width=\columnwidth]{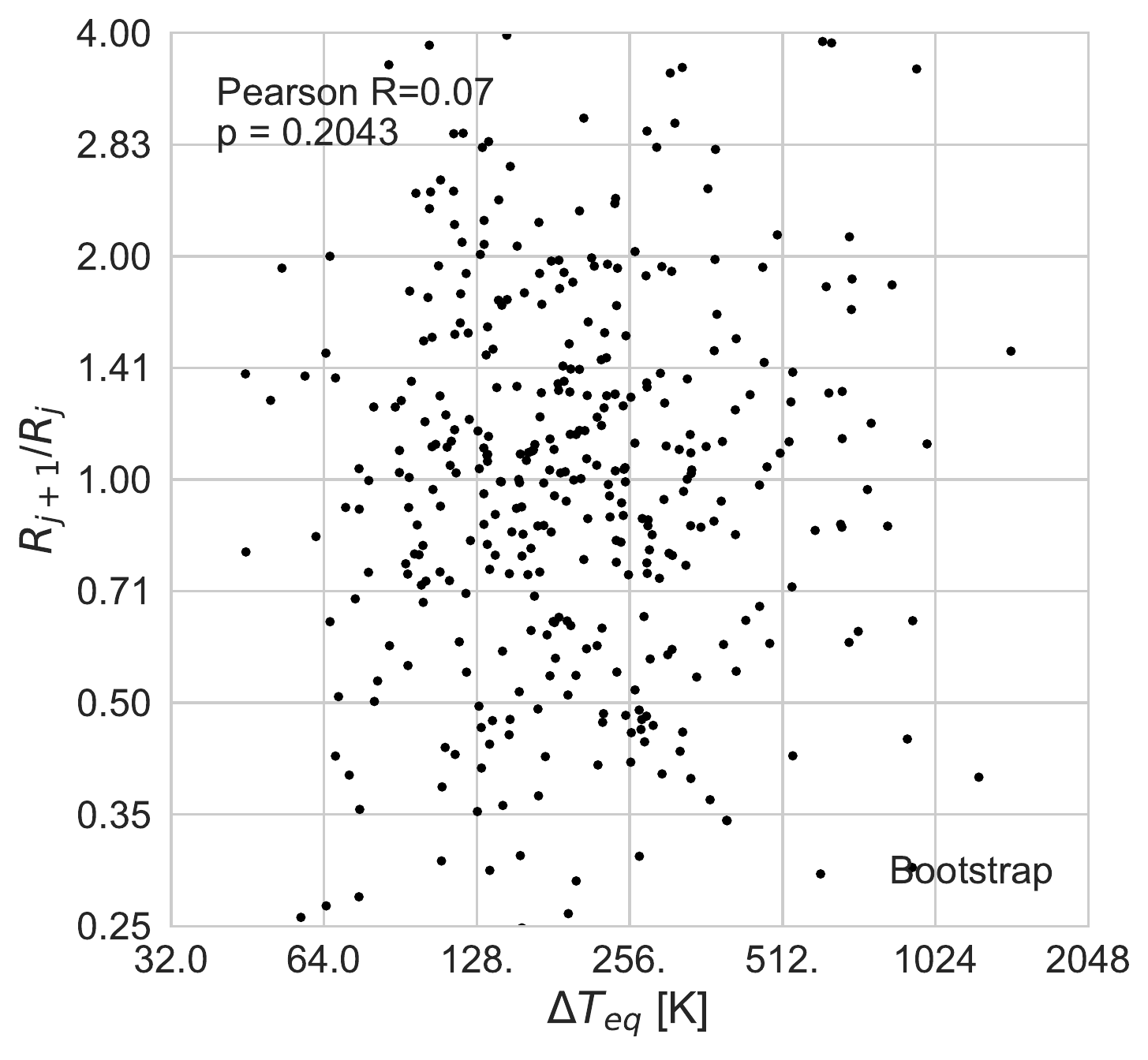}
   \caption{The ratio of adjacent planet sizes vs. the difference in their equilibrium temperatures in one of 1000 bootstrap trials.  The lack of correlation in the bootstrap trial is different from the modest correlation between equilibrium temperature different and planet size ratio (Figure \ref{fig:teq-corr}).}
   \label{fig:teq-bootstrap}
\end{figure}

Is the correlation between the planet size ratio and temperature difference due to astrophysical processes such as photo-evaporation, or a detection bias?  Repeating the bootstrap procedure of Section \ref{sec:sizes}, we find that bootstrap trials do not reproduce the observed distribution (Figure \ref{fig:teq-bootstrap} shows one example).  Rather, the bootstrap trials generate detectable planet pairs that populate the lower right corner of the plot (pairs in which the inner planet is larger, even when the temperature difference is large), demonstrating that such planets would have been detectable.  Thus, the size ratio vs. temperature difference trend is likely due to astrophysics, although the significance is marginal.  Determining whether photo-evaporation or some other physics is underlying the correlation will require further investigation and is beyond the scope of this paper.

\citet{Ciardi2013} also found that, after correcting for detectability, the outer planet is larger than the inner planet in 68\% of pairs, but this was only true for planets larger than 3~\rearth.  Restricting their sample to planets smaller than 3~\rearth, they found no preference for smaller inner planets, whereas we do (67\% of outer planets are larger in pairs where both planets are smaller than 3~\rearth; see Figure \ref{fig:rpcorr}).  The difference in their analysis and ours likely results from our larger sample. 


\section{Summary and Discussion}
\label{sec:conclusion}
\label{sec:summary}
In this paper, we have found the following observational results among the \Kepler\ systems with multiple transiting planets:
\begin{enumerate}
\item In a given multiplanet system, each planet's size is strongly correlated with the size of its neighbor.
\item In a given system with three or more transiting planets, the ratio of orbital periods between two adjacent planets is correlated with the orbital period ratios of other adjacent planets in that system.
\item There is a correlation between planet size and planet spacing: smaller planets tend to have smaller orbital period ratios.
\item Converting planet radii to estimated masses, we find that planets tend to be about 20 mutual Hill radii apart.  Planets are rarely closer together than 10 mutual Hill radii.  Higher multiplicity systems are more packed than lower multiplicity systems.
\item Planets spanning 0.5 to 4~\rearth\ have orbital period ratios as small as 1.2, but not smaller.  This corresponds to a mutual Hill radius of $>16$ for the smallest planets ($\rpl < 1~\rearth$).
\item In 65\% of planet pairs, the outer planet is larger than the inner planet.  The ratio of the outer planet to inner planet size is correlated with the temperature difference between the inner and outer planet.
\end{enumerate}

How do the observed patterns in the CKS multis relate to theories of planet formation?  The correlation between planet size and planet spacing---at least for the close planetary pairs---suggests that dynamics play a key role in the final planet sizes and/or the final planet spacings.  In particular, how do the patterns in the CKS multis relate to theories of \textit{in situ} formation vs. disk migration?

\subsection{\textit{In situ} formation}
\citet{Lissauer1993} noted, ``The self-limiting nature of runaway growth strongly implies that massive protoplanets form at regular intervals in semimajor axis.''  In short, this is because protoplanets grow until they have accumulated all of the available material in their feeding zone\footnote{an annulus centered on the star in which the velocity difference between the protoplanet and a swarm of planetesimals is sufficiently small that the gravitational force of the protoplanet wins over competing forces on the planetesimals \citep{Lissauer1993,Goldreich2004}}.  Likewise, \citet{Kokubo1998} noted, ``Protoplanets with the same order masses...[are] the inevitable outcome of planetary accretion.''  If oligarchs that formed via runaway growth remained undisturbed since their formation, they should still be at regular intervals today, and should still have similar masses.  That the CKS multis are similarly sized planets at regular intervals might indicate that they are aged oligarchs that suffered relatively few mass-doubling giant impacts, compared to the solar system terrestrial planets.  For comparison, the standard deviation of $R_{j+1}/R_j$ among the CKS multis is \rpratstd, whereas among the solar system terrestrial planets it is 1.01.  Future work that reproduces the typical intra-system variance in planet size and spacing via a detailed dynamical model might be revealing about the prevalence of giant impacts in the CKS multis.

Protoplanetary formation theory does not generally include the acquisition of gas, which is an important factor in the final planet size, since just a few percent hydrogen by mass can double a planet's radius.  If the protoplanets of similar masses are forming at the same time, as long as the gas fraction does not vary by many orders of magnitude between the innermost and outermost planet, it is plausible that the planets acquire similar amounts of gas, which would explain why they grow to be the same size.  \citet{Millholland2017} find that planets in the same system tend to have similar masses, strengthening the evidence that similarly sized planets are remnants of planet formation. 

\subsection{Disk Migration}
On the other hand, perhaps the CKS multis experienced Type I disk migration that brought them to their present locations.  One criterion for this formation mechanism is that the majority of the planets are not caught in mean motion resonances \citep{Fabrycky2014}.  To escape resonant capture, the planets would need low masses (compared to their stars) and sufficiently high eccentricities \citep{Pan2017}.  However, because the Type I migration rate scales with planet mass, the planet masses might need to be finely tuned to reproduce the observed correlated spacing of planets, or would need to escape from resonance after migrating in a locked configuration.  Furthermore, for many of the \Kepler\ systems, the tight spacing ($\Delta < 20$) requires low eccentricities for dynamical stability \citep{Fang2012,Pu&Wu2015, Petrovich2015,Dawson2016}, which places an upper bound on plausible eccentricities achieved during migration.

One particularly intriguing population for disentangling the histories of \textit{in situ} formation vs. disk migration is the very small planets ($\rpl < 1~\rearth$), which have orbital period ratios of 1.2, but wide separations ($\gtrsim20$) in terms of mutual Hill radii.  That the planets are not closer together than a period ratio of 1.2, even though Hill stability allows it, is probably evidence that chaos and eventual Lagrange instability dominates at $P_{j+1}/P_j < 1.2$.  Although these planets could, in theory, have orbital period ratios larger than 1.2, all of the pairs detected so far have orbital period ratios clustered near the wall at 1.2.  What can the absence of sub-Earths at larger orbital period separations (and larger mutual Hill radii) tell us about formation theory?  Either (1) sub-Earths form near the stability limit, or (2) migration tends to bring sub-Earth sized planets to the stability limit and park them at that limit.

\subsection{Kepler and the Solar System}
In the bottom panel of Figure \ref{fig:hill_sep}, the cumulative distribution function for the solar system (dotted line) traces the distribution for the CKS multis.  It is difficult to calculate the significance of the similarity because the solar system has only 9 planets (including Pluto).  The similarity in the mutual Hill radii between the CKS multis and the solar system is striking because the \Kepler\ planetary systems are often distinguished from the solar system with the phrase ``dynamically packed.''  However, Figure \ref{fig:hill_sep} underscores that in a dynamical sense, their orbital separations are very similar.  The orbital separations of the \Kepler\ planets in units of a.u. are small compared to the solar system, but this is not so in units of mutual Hill radii.\footnote{\citet{Malhotra2015} and others have also made this point.}

Nonetheless, the \textit{inner} solar system (our terrestrial planets) are quite far apart in mutual Hill radii, unlike the \Kepler\ systems.  The top panel of Figure \ref{fig:hill_sep} shows that Venus and Earth, Earth and Mars, and Mars and Mercury are all $> 20 $ mutuall Hill radii apart (Mercury and Venus are more than 60 mutual HIll radii apart).  Also, the orbital period ratios between all these planet pairs are larger than 1.2.  Although our solar system has an overall resemblance, in terms of dynamical packing, to the CKS multis, the amount of space between the terrestrial planets is rare among the \Kepler\ planets.  

Perhaps the wide spacing of planets in our inner solar system is due to the influence of Jupiter \citep{Chambers2001, Levison2003, Gomes2005}.  A more rigorous search for giant companions to the \Kepler\ multi-planet systems is necessary to better contextualize our solar system among other multi-planet systems.

\subsection{The future}
Measurements of the masses and eccentricities in systems of regularly spaced planets, especially in systems where photo-evaporation has played at most a minor role, will test the predictions of \textit{in situ} formation models and Type I migration models.  Obtaining accurate planet multiplicity, planet masses, and planet orbital dynamics in such systems might elucidate how the majority of the \Kepler\ sub-Neptune sized planets formed.

Since the TESS primary mission is expected to obtain at most a year of continuous photometry \citep[in the continuous viewing zones][]{Ricker2015}, it will not be as sensitive to long-period planets in multi-planet systems as \Kepler\ was.  Additional planet searches using radial velocity data, transit follow-up, astrometry from Gaia, and direct imaging with a small inner working angle (such as from WFIRST with a starshade) will help extend our sensitivity to as many planets as possible in multi-planet systems.  Such observations are necessary to further test the predictions of planet formation theories and to understand if our solar system is common or rare.

\facilities{Keck:I (HIRES),  Kepler}

\acknowledgments{
The CKS project was conceived, planned, and initiated by AWH, GWM, JAJ, HTI, and TDM. 
Keck time for the CKS multis project was acquired by GWM, with assistance from LMW.
The observations were coordinated by HTI and AWH and carried out by AWH, HTI, GWM, JAJ, TDM, BJF, LMW, EAP, ES, and LAH. 
AWH secured CKS project funding. 
SpecMatch was developed and run by EAP and SME@XSEDE was developed and run by LH and PAC. 
Downstream data products were developed by EAP, HTI, and BJF.
Results from the two pipelines were consolidated and the integrity of the parameters were 
verified by AWH, HTI, EAP, GWM, with assistance from BJF, LMW, ES, LAH, and IJMC.
EAP computed derived planetary and stellar properties with assistance from BJF.  
LMW performed the analysis in this paper, with assistance from GWM, JW, and AC.
This manuscript was largely written by LMW with assistance from JW and GWM.

We thank John Asher Johnson, a PI and originator of the magnitude-limited CKS survey.  Conversations with Rebekah Dawson, Daniel Fabrycky, Eve Lee, Doug Lin, Jack Lissauer, Jason Rowe, and Scott Tremaine influenced this paper.
The results presented herein were made possible by observations at the W.\ M.\ Keck Observatory, which is operated as a scientific partnership among the California Institute of Technology, the University of California, and NASA. We are grateful to the time assignment committees of the University of Hawaii, the University of California, the California Institute of Technology, and NASA for their generous allocations of observing time that enabled this large project.
Kepler was competitively selected as the tenth NASA Discovery mission. Funding for this mission is provided by the NASA Science Mission Directorate.  
L.\ M.\ W.\ acknowledges support from Gloria and Ken Levy and from the Trottier Family Foundation.
E.\ A.\ P.\ acknowledges support from Hubble Fellowship grant HST-HF2-51365.001-A awarded by the Space Telescope Science Institute, which is operated by the Association of Universities for Research in Astronomy, Inc. for NASA under contract NAS 5-26555. 
B.\ J.\ F.\ acknowledges National Science Foundation Graduate Research Fellowship under Grant No. 2014184874. 
A.\ W.\ H.\ acknowledges NASA grant NNX12AJ23G.  
T.\ D.\ M.\ acknowledges NASA grant NNX14AE11G.
P.\ A.\ C.\ acknowledges National Science Foundation grant AST-1109612.
L.\ H.\ acknowledges National Science Foundation grant AST-1009810.
E.\ S.\ is supported by a post-graduate scholarship from the Natural Sciences and Engineering Research Council of Canada.
Finally, the authors wish to recognize and acknowledge the very significant cultural role and reverence that the summit of Maunakea has always had within the indigenous Hawaiian community.  We are most fortunate to have the opportunity to conduct observations from this mountain.}

\bibliography{references}
\bibliographystyle{apj}

\begin{deluxetable*}{cccccccccccccc}
\tablecaption{CKS Multis Properties}
\label{tab:multis}
\tabletypesize{\footnotesize}
\tablenum{1}
\tablehead{\colhead{Star} & \colhead{KOI} & \colhead{Kepler Name}& \colhead{$M_\star$} & \colhead{$R_\star$} & \colhead{CDPP$_\mathrm{6hr}$} & \colhead{$b$} & \colhead{Period} & \colhead{\rpl} & \colhead{$\sigma\rpl$} & \colhead{$T_\mathrm{eq}$} & \colhead{$\rpl$ avg.} & \colhead{$P_{j+1}/P_j$} & \colhead{$\Delta_\mathrm{RH}$} \\ 
\colhead{}  & \colhead{} & \colhead{} & \colhead{(\msun)} & \colhead{(\rsun)} & \colhead{} & \colhead{} & \colhead{(days)} & \colhead{(\rearth)} & \colhead{(\rearth)} & \colhead{(K)} & \colhead{(\rearth)} & \colhead{} & \colhead{} } 
\startdata
K00041  & K00041.02 & Kepler-100 b & 1.107 & 1.55 & 23.33 & 0.51 & 6.88 & 1.35& 0.23 & 1186 & 0.0 & 0.0 & 0.0 \\
K00041  & K00041.03 & Kepler-100 d & 1.107 & 1.55 & 23.33 & 0.59 & 35.33 & 1.54 & 0.31 & 687 & 1.95 & 2.75 & 31.23 \\
K00041  & K00041.01 & Kepler-100 c & 1.107 & 1.55 & 23.33 & 0.58 & 12.81 & 2.37 & 0.34 & 965 & 1.86& 1.86 & 20.15 \\
K00046  & K00046.01 & Kepler-101 b & 1.156 & 1.58 & 54.61 & 0.03 & 3.48 & 5.69 & 0.73 & 1443 & 0.0 & 0.0 & 0.0 \\
K00046  & K00046.02 & Kepler-101 c & 1.156 & 1.58 & 54.61 & 0.41 & 6.02 & 1.18 & 0.16 & 1203 & 3.43 & 1.72 & 12.91 \\
\enddata
\tablecomments{$\rpl$ avg., $P_{j+1}/P_j$, and $\Delta_\mathrm{RH}$ are parameters that relate a pair of adjacent planets.  In this table, each value is listed with the\\ \textit{outer} planet of the pair.\\
Table \ref{tab:multis} is available in its entirety in machine-readable format. A portion is shown here for guidance regarding its form and content.}
\end{deluxetable*}

\end{document}